# Calibration of MAJIS (Moons And Jupiter Imaging Spectrometer): III. Spectral Calibration

Paolo Haffoud[1,a], François Poulet[1], Mathieu Vincendon[1], Gianrico Filacchione[2], Alessandra Barbis[3], Pierre Guiot[1], Benoit Lecomte[1], Yves Langevin[1], Giuseppe Piccioni[2], Cydalise Dumesnil[1], Sébastien Rodriguez[4], John Carter[1], Stefani Stefania[2], Leonardo Tommasi[3], Federico Tosi[2], Cédric Pilorget[1]

[1]*Institut d'Astrophysique Spatial, CNRS-Université Paris-Saclay, Orsay, 91400, France*

[2]*Istituto di Astrofisica e Planetologia Spaziali, Istituto Nazionale di Astrofisica, Rome, 00133, Italy*

[3]*Leonardo Company, via delle Officine Galileo, 1, Campi Bisenzio, 50013, Italy*

[4]*Institut de Physique du Globe de Paris, CNRS-Université Paris-Cité, Paris, 75005, France*

a) *The author to whom correspondence may be addressed: paolo.haffoud@universite-paris-saclay.fr*

The Moons And Jupiter Imaging Spectrometer (MAJIS) is the visible and near-infrared imaging spectrometer onboard ESA's Jupiter Icy Moons Explorer (JUICE) mission. Before its integration into the spacecraft, the instrument undergoes an extensive ground calibration to establish its baseline performances. This process prepares the imaging spectrometer for flight operations by characterizing the behavior of the instrument under various operative conditions and uncovering instrumental distortions that may depend on instrumental commands. Two steps of the on-ground calibration campaigns were held at the instrument level to produce the data. Additional in-flight measurements have recently been obtained after launch during the Near-Earth Commissioning Phase. In this article, we present the analyses of these datasets, focusing on the characterization of the spectral performances. First, we describe and analyze the spectral calibration datasets obtained using both monochromatic sources and polychromatic sources coupled with solid and gas samples. Then, we derive the spectral sampling and the spectral response function over the entire field of view. These spectral characteristics are quantified for various operational parameters of MAJIS, such as temperature and spectral binning. The derived on-ground performances are then compared with in-flight measurements obtained after launch and presented in the framework of the MAJIS performance requirements.

## I. INTRODUCTION

The Jupiter Icy Moons Explorer mission (JUICE) was launched in April 2023 towards the Jovian system, which will be reached in July 2031. JUICE will first tour the Jovian system for 3.5 years, during which 35 flybys of the 3 Galilean Moons (Callisto, Ganymede, and Europa) will be performed before entering orbit around Ganymede at the end of 2034 for about nine months[1,2,3]. The Moons And Jupiter Imaging Spectrometer (MAJIS) will acquire hyperspectral data cubes of Jupiter's atmosphere and the surfaces and exospheres of the three icy moons to determine their composition and identify various chemical species[4]. Of particular interest are the study of Ganymede's ice-shell and exosphere, the investigation of the chemistry of surface material and active processes on Europa, and the monitoring of the Jupiter atmosphere with unprecedented temporal coverage and spatial and spectral resolutions in the near-infrared wavelength range. The MAJIS science goals and science operations and whole instrument design have previously been described in details[4]. We present below the necessary elements which are relevant for spectral calibration.

### A. MAJIS characteristics relevant for spectral calibration

The optical unit of MAJIS is made of two co-aligned spectral channels[4]. One is a VISible to Near-InfraRed (VISNIR) channel, and one is an InfraRed (IR) channel. The Optical Head (OH) of MAJIS incorporates a Three Mirror Anastigmat (TMA) telescope shared between the two channels. A slit and a collimator, common to both channels, control light input and ensure parallel propagation towards a dichroic filter. This filter splits the light into the two channels. Each channel is equipped with a grating following a quasi-linear dispersion law and a focusing lens for spectral analysis. Both channels collect spectral images of a common slit on their Focal Plane Arrays (FPAs). Due to optical aberrations in the OH of the imaging spectrometer, a spectral shift (smile effect) of the sensor over its entire Field Of View (FOV) is present. (e.g., imaging a homogeneous target over the full FOV can result in misaligned spectra for different FOV positions). The FPAs are 1024×1024 pixels H1RG Teledyne detectors with an 18×18 µm² pitch. The row and column directions correspond to the spectral and spatial dimensions, respectively. Any pixel will be referred to as "spectel" hereafter when its spectral properties are considered. During the nominal scientific operations, these 18 µm pixels are binned by two in the spatial direction or more (spatial and/or spectral direction) depending on the selected spectral or spatial binning commanded by telemetry (see below). Thus, in the following sections of this article, "nominal pixel" will refer to four physical pixels binned in a two-by-two square (i.e., 36 µm wide). The instrument's FOV is projected onto the two FPAs across 800 contiguous rows of 18 µm pixels; the wavelengths are dispersed across 1016 columns of 18 µm spectels. After nominal binning, this results in a nominal frame acquisition of size 400×508 pixels for each channel.

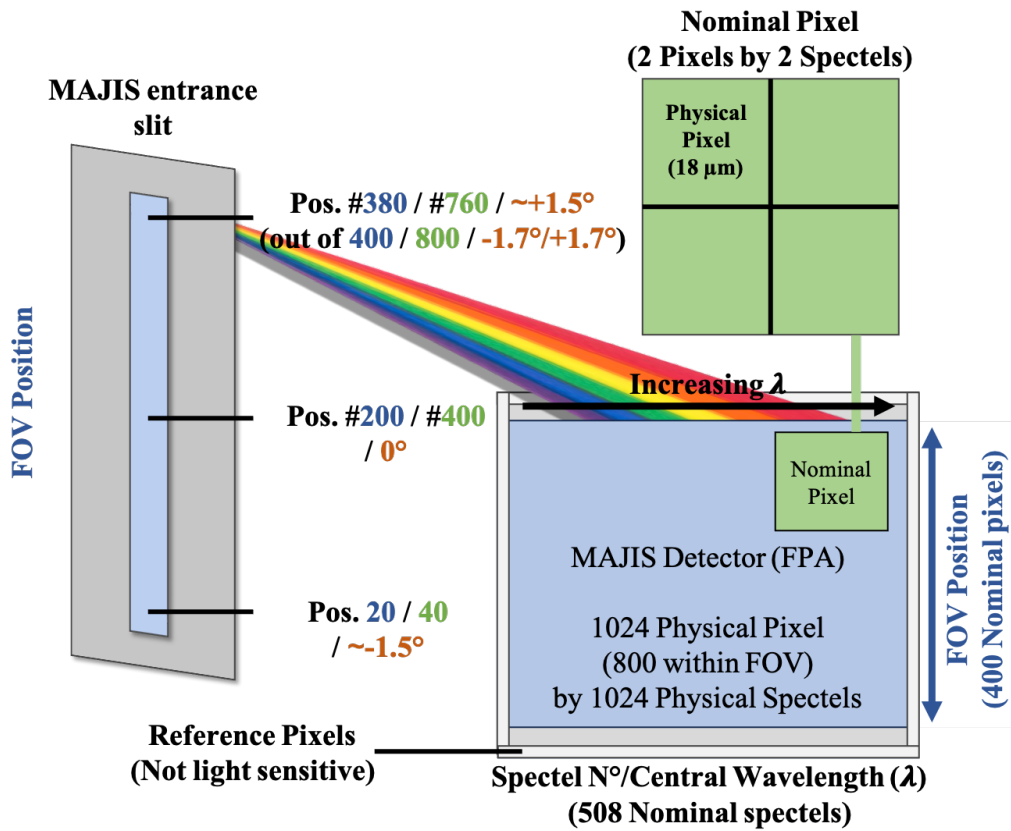

Fig. 1. Schematic view summarizing the nomenclatures related to the FOV and spectel positions of MAJIS. Positions (in pixel) inside of the FOV are exemplified. The positions along the FOV are given in terms of nominal pixels (blue), physical pixels (green), and angular position (red).

MAJIS has the capability to perform spectral oversampling and binning of its nominal pixels over specific spectral ranges (referred to as spectral bands)[4]. Oversampled pixels (18 μm) are not binned by two over the spectral dimension. The binning ×2 and binning ×4 modes allow the physical pixels (18 μm) to be grouped by 4 and 8, respectively. Up to 16 spectral bands with different oversampling/binning modes can be defined per channel during one observation. A specific type of oversampling/binning is implemented for each spectral range. Oversampling can increase the number of spectels per frame to a maximum of 640 pixels instead of the nominal 508. Higher spectral sampling rates will provide more detailed spectral information. However, this must be balanced with the collected data quality regarding the Signal-to-Noise Ratio (SNR) and the data volume of the observations.

MAJIS also has an acquisition mode called *No Process*. This mode transmits the 1024 by 1024 raw images. This mode will be infrequently utilized during science operations due to its large data size and inefficient use of memory space/data bandwidth (> 200 by 1024 pixels provide no usable data). However, during ground calibration, these disadvantages are nullified. Furthermore, the data provided by this mode can easily be numerically post-processed (e.g., numerical binning) to simulate any acquisition settings used in flight and derive its associated spectral calibration performances. Therefore, the vast majority of ground calibration measurement series was performed using the *No Process* mode. This article will describe the spectral calibration performances for the nominal, oversampling, binning ×2, and binning ×4 modes using data acquired with the *No Process* acquisitions.

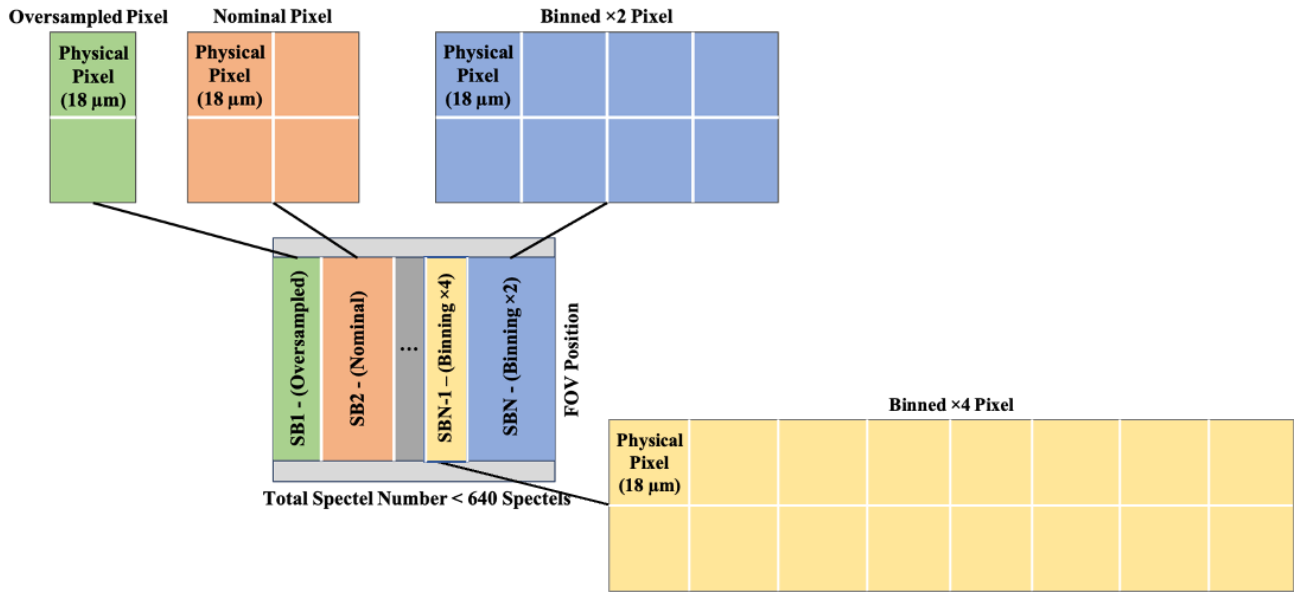

Fig. 2. Example of a MAJIS frame summarizing its oversampling/binning modes capabilities. The resulting frame (up to 640 spectels) can combine up to 16 spectral bands with a specific spectral mode defined by TC (TeleCommand) for each spectral band.

A scanning mirror at the entrance of the telescope provides the capability to change the instrument's pointing through a 4° across-slit optical angle centered around the boresight. This scanning mirror can also point to the Internal Calibration Unit (ICU) light sources[4] that provide relevant data to check the fly performances of the instrument.

### B. Spectral calibration requirements and goals

Advanced hyperspectral imagery, as MAJIS does, implies a complex instrumental design. This results in several instrumental effects within the datasets to be characterized. To properly extract and derive the relevant scientific information, an in-depth ground characterization of the instrumental effects of MAJIS was performed from subsystems to integrated system[4,5].

Imaging spectrometers are susceptible to optical distortions caused by optical aberrations and misalignments among optical elements. Even if these can be reduced by optimizing the optical layout and utilizing a mechanical mount to keep the optical elements in the correct positions, it is required to characterize the central wavelength position and the spectral response characterized by the Full Width at Half Maximum (FWHM) of each detector pixel after the mechanical and optical integration and the final alignment. The optical distortions can also result in a smile effect[6] (i.e., a wavelength drift along a detector row), and calibration variations can be sensitive to OH temperature (110-150K operative range for MAJIS). Therefore, the spectral calibration aims to assess these items with dedicated ground setups. The requirements for the MAJIS spectral calibration were driven by the MAJIS scientific performances summarized in Table I. However, this article provides a detailed analysis that is not limited to the validation of the requirements. As MAJIS oversampling capabilities can be used on any part of the spectral range (within the limitations mentioned in Section I.A), the strategy for the analysis was to focus on the instrumental performances of oversampled spectels. Then, the effects of the different types of binning on the performances are characterized for the different parameters. This allows us to deduce the performance of MAJIS in its various operational modes.

Table I. MAJIS spectral requirements for a nominal pixel (36-μm) in both directions (spectral and spatial, i.e., columns and rows on the detectors). These requirements shall be met across the entire operational thermal range of the instrument (110-150 K).

| Item | Metrics | Channel | MAJIS requirement |
|---|---|---|---|
| **Absolute Calibration** | Range | VISNIR | 500 - 2350 nm |
| | | IR | 2250 - 5540 nm |
| | Sampling | VISNIR | $3.65 \pm 0.25$ nm |
| | | IR | $6.50 \pm 0.60$ nm |
| **Spectral response** | FWHM | VISNIR | $\leq 5.5$ nm |
| | | IR | $\leq 10$ nm |
| **Spectral distortion along the FOV** | Smile | VISNIR | $\leq 3.90$ nm |
| | | IR | $\leq 7.10$ nm |

The ground calibration measurements dedicated to MAJIS were based on the legacy of several previous hyperspectral space imaging spectrometers[6,7,8,9]. The overall ground calibration strategy at sub-systems and instrument levels is described in Poulet et al.[4]. Specifically, the ground calibration at the instrument level was performed during two main campaigns. The first campaign occurred at LeonarDO Company (LDO) facilities (Firenze, Italy) to characterize the performances of the OH at three thermal configurations (nominal, hot, and cold cases) in using the focal plane acquisition pipeline. The following main campaign, held at the Institut d'Astrophysique Spatiale (IAS) in Orsay, was designed to provide the measurements required for the spectral (this article), spatial[10], and radiometric[11] calibration in using the nominal acquisition pipeline with the flight spare unit of the main electronic[5]. The launch of the JUICE mission was followed by the Near-Earth Commissioning Phase (NECP). During this commissioning, the MAJIS instrument acquired its first in-flight frames that could be used to monitor the consistency of the performances (including spectral ones) with the ground characterization.

Section II briefly describes the ground calibration setups and the datasets acquired with the different calibration sources (monochromatic scans, atmospheric measurements, internal calibration unit, sample, and mineral images). Section III introduces the methodology applied to the reduction of data before their analyses to retrieve the spectral performances of the instruments. Section IV presents the results of the calibration data analysis for each spectral requirement listed in Table 1, namely the spectral response, the spectral sampling, the spectral range, the smile, and the impact of the OH temperature on the spectral calibration. Using the internal calibration unit, a comparison between these performances and the first post-launch measurements is described in Section V. Finally, the spectral performances are summarized and discussed in Section VI.

## II. CALIBRATION SETUPS AND DATASETS

### A. Setups

Two calibration setups were used to establish the performances of MAJIS, including the spectral ones. The LDO setup aimed to provide a limited characterization of the performances of MAJIS OH prior to shipping for full calibration. The IAS setup was designed to calibrate the instrument thoroughly and validate that all performance requirements were met. The key features differentiating both calibration setups are summarized in Table II.

Table II. Main specific features of the calibration setups

|  | Source(s) | MAJIS OH temperatures explored (K) | FOV exploration | Spectral exploration | Controlled environment | Illuminating beam size |
|---|---|---|---|---|---|---|
| **LDO** | Monochromator (QTH / IR emitter) | 110, 130, 150 | Small | Small | OH inside TVC + Non-flight acquisition chain + Source mounted on a hexapod outside | Limited to entrance pupil |
| **IAS** | OP1: Monochromator | 126, 137 | Small | Small | OH mounted on a hexapod inside TVC | Covers the whole |

| | | | | | |
|---|---|---|---|---|---|
| | (QTH, blackbody) | | | | + Flight acquisition chain + Calibration setup optical paths in a flushed (N$_2$) environment | entrance baffle |
| | OP3: QTH + integrating sphere (extended source with gas) | | Wide | Wide | | |
| | OP5: QTH (Reflected light on solid sample) | | Wide | Wide | | |

For the spectral calibration, the LDO setup dataset was mainly used to provide additional monochromatic scans at wavelengths that were not explored during the full calibration. A few LDO scans covered a spectral range also encompassed by the IAS dataset. In addition to monochromatic scans, the IAS dataset provided atmospheric signatures spectra, solid samples spectra and ICU measurement spectra. The key elements of the calibration strategy were to use the monochromatic scans to calibrate the spectral response as well as to give some data points for the absolute calibration. The spectra acquired with non-monochromatic sources were used to derive several absolute calibration (and smile) data points based on the matching of signature spectral positions between MAJIS and reference values. These high-accuracy data points complimented parts of the spectral range not covered by the monochromatic scans. This strategy summary is illustrated in Table III.

1. LDO

This setup is described in more detail in a companion paper[10]. The first stage to collect data used here to derive the spectral calibration occurred at LDO in Italy[12]. The MAJIS OH was tested inside a Thermal Vacuum Chamber (TVC) at three OH operative temperatures (cold 110 K, nominal 130 K, and hot 150 K), illuminated by the Optical Ground Support Equipment (OGSE) input beam. The OH was fixed, and the optical beam was moved within the FOV by means of a hexapod. The input beam was adapted to cover only the entrance pupil (i.e., it did not cover the whole entrance baffle). The tests were driven by a master computer using automatic test procedure commands to control the OGSE devices and the MAJIS acquisitions during the measurements. The OGSE configuration used specifically for the spectral characterization includes an off-axis parabolic mirror to collimate the test beam, a source module including a monochromator with two sources (Quartz Tungsten Halogen lamp and IR emitter from Newport/Oriel) covering the spectral range between 750 and 5300 nm, target translation stages to scan the test slit along and across the OH slit, and an OGSE hexapod to orientate the collimated beam at various FOV values. Specifically, one or three positions inside the FOV were explored according to the thermal configurations of the OH.

## 2. IAS

The calibration setup used for IAS calibration is described in detail in a companion paper[5]. A previous configuration of this setup was used to calibrate the Spectrometer and Imaging for MPO BepiColombo Integrated Observatory SYStem (SIMBIO/SYS) instrument[13]. This setup was upgraded to meet all MAJIS calibration requirements. In particular, specific sources were implemented to fulfill the spectral calibration objectives previously mentioned. The OH and the main electronics were integrated into a thermal vacuum chamber. The OH itself was mounted on a hexapod housed in the thermovacuum chamber to allow the angular exploration of the FOV and the slit by the fixed beam from the calibration bench. The full entrance baffle of MAJIS was illuminated at IAS, while only the part of the first mirror corresponding to the entrance pupil was illuminated at LDO. The complete ground calibration setup has five optical paths, each having its own illumination source. They are referred to as OP# with numbering from 1 to 5. Channels OP1, OP3, and OP5 were used for spectral calibration. OP1 is the dedicated spectral source equipped with a monochromator and two sources to cover the spectral range of MAJIS (Table II). While OP3 was initially dedicated to the radiometric calibration of the VISNIR channel, this optical path was used after having stopped the nitrogen flushing to introduce atmospheric contamination inside the calibration setup, leading to telluric features in the MAJIS spectra. OP5 is the ground calibration setup assigned to the reflectance measurements of samples (minerals and Spectralon). In addition, spectral calibration data were acquired with the two sources of the ICU (also referred to as OP0). Below, we describe these sources in more detail.

The OP1 source provides the monochromatic illumination thanks to a monochromator illuminated by a stabilized Quartz Tungsten Halogen (QTH) source for the 300 nm to 2000 nm range and a thermally stabilized blackbody for the 2000 nm to 6000 nm range. The monochromatic source can be spatially adjusted thanks to the monochromator's output slit to ensure both the required flux and the spectral resolution. The parameters of the monochromatic source were optimized thanks to the monochromator's grating choice and output slit width: a constant spectral width of 0.7 and 1 nm for the VISNIR and IR channels were selected, respectively. Then, an optical relay system and a collimator redirect the collimated OP1 beam toward the window of the TVC facing the MAJIS telescope entrance. The optical relay system consists of lenses and mirrors positioned to guide light from the source to the collimator. Given the magnification ratio of the collimator, the monochromator's exit slit is imaged on about 20 nominal spatial pixels by MAJIS. Three FOV positions were explored to derive the spatial uniformity of the spectral response: nominal pixels 20, 200, and 380 out of the 400 introduced in Section I. These values correspond to the center of the FOV (position 200, boresight), bottom of the FOV (position 20, ~ -1,5°), and top of the FOV (position 380, ~ +1.5°).

The OP3 optical path was initially designed for the VISNIR radiometric calibration, but relevant data for assessing the spectral performances were also acquired with this setup. It consists of an integrating sphere connected to a QTH lamp as a light source to produce a flat field dedicated to the radiometric calibration of the VISNIR channel. An off-axis conical aluminum mirror collimates the sphere output. The output of the integrating sphere is large enough to cover the entire FOV of MAJIS. The source and the collimating mirror are placed in an $N_2$-flushed extension of the calibration optical tank directly in front of the $CaF_2$ window of the TVC, limiting the 293 K thermal contribution and potential $H_2O$ signatures while offering a wide range of light intensity controlled by the current of the QTH and the position of a diaphragm placed at the interface of the QTH and the entrance port of the integrating sphere. The integrating sphere of OP3 thus provides a homogeneous beam on its output port, with calibrated probes controlling the flux level of the QTH. For the spectral calibration, this optical path was used without $N_2$ flushing to obtain transmission spectra through the ambient air, which hence exhibited water vapor spectral features. The $N_2$ flush was also replaced by a $CO_2$ flush to obtain measurements of the $CO_2$ absorptions (Section II.C.2.a).

The OP5 path uses another QTH source to illuminate solid samples in reflection. The reflected flux is then directed toward the same collimator used for OP1 using dedicated optics. MAJIS then images the samples. Solid samples relevant for the spectral calibration are 1) a Wavelength Calibration Standard (WCS) with multiple absorption features over the VISNIR, and 2) a calcite pressed-pellet with complementary absorption bands. Due to thermal constraints, data are limited to wavelengths < 3000 nm (samples are not cooled).

The ICU sub-system is mounted inside the entrance baffle of MAJIS. It comprises a VISNIR incandescent lamp and an IR MEMS source illuminating a common diffusive coating[14]. The signal of the two sources is collected by rotating the internal scan mirror in correspondence with the ICU boresight. Each source is equipped with a spectral filter to provide reference spectral signatures on the MAJIS signals. Specifically, the VISNIR source uses a didymium filter, and the IR source a polystyrene filter[12]. The primary purpose of the ICU is to monitor the evolution of the spectral performances of the instrument during in-flight operations. ICU measurements were acquired during the LDO and IAS campaigns to provide reference data for the ground and pre-flight performances of the instrument.

3. **Thermal configuration**

Table IV, Table V, and
Table VI indicate that various thermal steady-state configurations were tested during the LDO and IAS calibration. At the LDO premise, the cryogenic temperatures were achieved through a dedicated thermal control system. Three thermal cases were considered: OH structure stabilized at 110, 130, and 150 K, corresponding to the lower, nominal, and upper operative cases, respectively. A more restricted data set was obtained for the lower and upper operative cases since these cases are very unlikely during the science operation. At IAS, the thermal control of the IR detector in the TVC was managed with a cryocooler, and liquid nitrogen was used to cool the OH and the VISNIR FPA. The thermal exploration has been selected to cover two operative configurations expected to be representative of most measurements during the science operations at Jupiter: 126 K (nominal thermal case), where most of the sequences were completed, and 137 K (hot case), where part of the sequences was performed again.

### B. Data format

As shown in Table I, the requirements are specified for nominal pixels (36-µm pitch). Still, the acquisitions (and consequently the analysis) are performed using the specific readout procedure *No Process* that allows reading the full frames (1024 × 1024) at the level of the physical pixels (18-µm pitch) in both dimensions (see section I.A). This mode has the advantage of keeping the full detector resolution, which was required to evaluate MAJIS performance in the oversampling (i.e., un-binned) optional mode. Binning can be easily implemented after data acquisition with the same algorithm used by the onboard software. Therefore, unless specified otherwise, the analyses and the results of the spectral calibration are given in terms of physical spectels corresponding to the oversampling mode, and the effects of binning are discussed in a dedicated paragraph (Section IV.C).

Several measurements were acquired using the capability of reading out only part of the 1024 rows along the spatial direction thanks to the MAJIS capability of spatial windowing. This mode results in a shorter repetition time, and it is useful when the source covers only a part of the FOV or when a shorter repetition time is required (for instance, to avoid saturation). The use of this mode has no impact on the derivation of the spectral performances.

### C. Data description

The summary of the data sets acquired at LDO and IAS premises is presented in Table III. Table IV, Table V, and Table VI provide additional details on the measurements (wavelengths, position on the FOV, OH temperature…) for each source. Because of the planning constraints, the parameters were adjusted during the calibration campaigns.

Table III. Overview of data sets used for the data analysis. The monochromatic data acquired at LDO were processed similarly and used for the same purposes as the IAS OP1 data. This data will also be referred to as OP1 in the article.

| **Measurement type** | Internal calibration unit (see Table VI) | Monochromatic light (see Table IV and Table V) | Gas sample in transmission (see Table VI) | Solid sample in reflectance (see Table VI) |
|---|---|---|---|---|
| **Spectral response** | | LDO / IAS datasets | | |
| **Absolute spectral calibration** | IAS dataset | LDO / IAS datasets | IAS dataset | IAS dataset |
| **Optical source reference number** | OP0 | OP1 | OP3 | OP5 |

1. **Monochromator scans**

The monochromatic measurements with IAS and LDO OP1 setups consist of several series of frames captured while receiving a monochromatic beam with a full width at half maximum several times smaller than the expected spectral response FWHM. For each position of the monochromator scan, one or more frames were captured in order to determine the correlation between the selected wavelength and the observed signal as a function of the position along the MAJIS spectral axis. Moreover, each scan across the wavelength range of the detector aims to characterize the spectral response of the illuminated spectels and determine the absolute central wavelength. A tradeoff taking into consideration the signal-to-noise of the data, the capability of setups, and the required number of measurements to derive the spectral response of a given spectel with acceptable accuracy led to performing a spectral scan with a scan step of ~1/5 of the expected spectral sampling of the instrument (i.e., 0.7 nm (VISNIR channel) or 1 nm (IR channel)) for the IAS data and 0.5 nm (VISNIR channel) or 1 nm (IR channel) for LDO data[5]. An example of a frame acquired with the IAS OP1 setup is shown in Fig. 3 (A). The accuracy of the measurements depends on the accuracy of the monochromatic sources. The absolute accuracy is 1 nm (resp. 0.05 nm) for both IAS and LDO setups (relative accuracy). The monochromator spots were around 20 physical pixels wide (14 pixels for the IAS setup with the VISNIR channel) in the along-slit direction (which is equivalent to 10 nominal pixels out of the 400 nominal FOV pixels). Fig. 3 (B) displays the spectra derived from sequential measurements during a monochromatic scan. The generation of these spectra can be summarized in two steps. First, the quasi-monochromatic signal from the monochromator is convolved with the spectral response of MAJIS, creating a combined spectrum. This combined spectrum is then sampled by MAJIS to produce the final spectra shown.

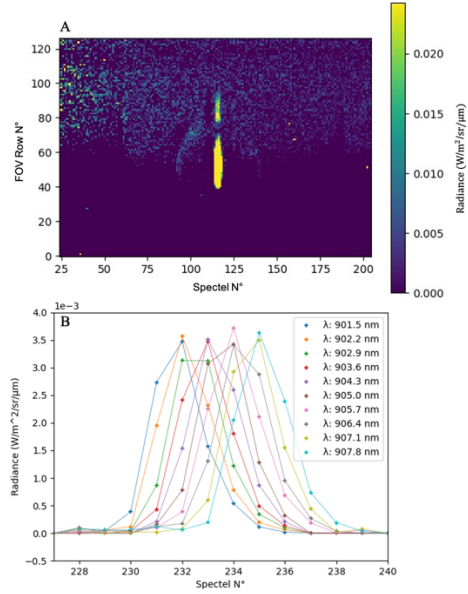

Fig. 3. Panel A: Illustration of a monochromatic acquisition around 700 nm at the FOV position 200 (nominal). The main central spot (between rows ~38 to ~70) of the monochromator is spectrally and radiometrically uniform for about 20 physical pixels along-slit. Panel B: Example of spectra of 10 median images (out of 51) from a monochromatic scan around 900 nm at the FOV position 200.

Due to the setup differences between IAS and LDO (see Section II.A.), the OP1 light source only covered the entrance pupil of the OH at LDO facilities, whereas the full entrance baffle was covered by a light source from the setup at IAS. This was not expected to have an impact on the resulting acquisitions. However, due to issues from either the design or the assembly of the instrument, residual stray light effects can be seen (mainly on the VISNIR channel). Therefore, light originating from outside the entrance pupil can be seen on the acquisitions. These imperfections hint at the possibility that the actual performances of the instrument will differ from the performances when only the entrance pupil receives light, especially for the VISNIR channel.

Table IV and Table V describe the datasets acquired during IAS calibration and LDO characterization. The spectral and spatial coverage is also illustrated in Fig. 4. As seen on this figure, the number of spectral intervals and positions inside the FOV that were explored was restricted because of the planning constraint. Achieving a good signal-to-noise for wavelengths beyond 3.6 µm was also very challenging due to the thermal background and the low level of the source flux. The mitigation of these limitations was to complement monochromatic measurements with polychromatic measurements over solid samples and gas, as described in the following section.

Table IV. Monochromator scan measurements series from IAS OP1

| Channel | Wavelength Range (nm) | OH / VISNIR FPA (and IR FPA) Temperature (K) | MAJIS Slit Position | Scan Step (nm) | Monochromator spectral bandwidth (FWHM) (nm) |
|---|---|---|---|---|---|
| VISNIR | 550-585 | 126 (88) | 200 | 0.7 | 0.7 |
|  | 900-935 | 126 (88), 137 (97) | 20, 200, 380 | 0.7 | 0.7 |
|  | 1400-1435 | 126 (88), 137 (97) | 20, 200, 380 | 0.7 | 0.7 |
|  | 1900-1935 | 126 (88), 137 (97) | 20, 200, 380 | 0.7 | 0.7 |
| IR | 2600-2650 | 126 (88), 137 (97) | 20, 200, 380 | 1 | 1.3 |

Table V. Monochromator scan measurements series from LDO OP1

| Channel | Wavelength Range (nm) | OH / VISNIR FPA (and IR FPA) Temperature (K) | MAJIS Slit Position | Scan Step (nm) | Monochromator spectral bandwidth (FWHM) (nm) |
|---|---|---|---|---|---|
| VISNIR | 687.5-712.5 | 130 (91) | 200 | 0.5 | 0.5 |
|  | 1387.5-1412.5 | 130 (91) | 20, 200, 380 | 0.5 | 0.55 |
|  | 2087.5-2112.5 | 130 (91) | 20, 200, 380 | 0.5 | 0.55 |
| IR | 2275-2325 | 130 (91) | 20, 200, 380 | 1 | 3.15 |

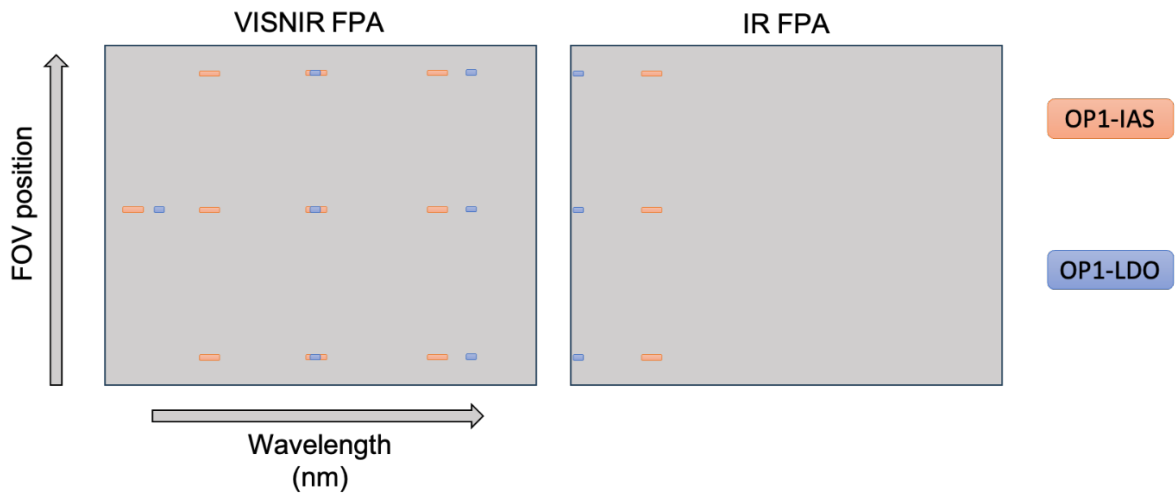

Fig. 4. Schematic view of the coverage obtained from the monochromatic scan dataset acquired at IAS and LDO. The spectral positions of the scans are approximate.

2. **Additional spectra measurements**

Additional MAJIS calibration measurements useful to the spectral calibration are described in this section. They are separated into three subsections corresponding to three different OPs and are described in Table VI.

Table VI. Atmospheric measurements series main parameters

| Measurement type | Channel | OH Temperature (K) | Position |
|---|---|---|---|
| **Atmospheric** | VISNIR / IR | 126, 137 | Full FOV |
| **Internal Calibration Unit** | VISNIR / IR | 126, 137 | Full FOV |
| **Solid Samples** | VISNIR / IR | 126, 137 | 20, 200, 380 |

*a. Atmospheric absorption*

OP3[5] was purposely used after the flushing was stopped so that atmospheric gases such as water vapor and carbon dioxide could be identified in the MAJIS frames. In addition, $CO_2$ gas was flushed in OP3 to increase its concentration along its 1-m long optical path so as to increase the atmospheric signatures beyond 3 μm. The positions of the telluric signatures provide unique reference points for the absolute calibration of MAJIS. The High Resolution Transmission (HITRAN)[15] model was used to calibrate the absolute wavelength positions of the various signatures observed with the measurements. Note, however, that the integrating sphere of OP3 was adapted to characterize the instrument in the VISNIR range (see Section II.A.2). This limits the capability to identify atmospheric absorption bands in the longest IR wavelengths where the flux is dominated by the thermal contribution of the optical elements that are at ambient temperatures. Conversely, these acquisitions have the advantage of covering the full FOV in one acquisition.

### b. Internal Calibration Unit

The ICU acquisitions (OP0) provided additional reference measurements for the absolute calibration thanks to the didymium and polystyrene signatures, and in this respect, they were used for the on-ground spectral calibration. Samples of polystyrene similar to the ICU material were spectrally characterized at the Istituto di Astrofisica e Planetologia Spaziali (IAPS, Rome) facilities in the same temperature range to allow comparison with MAJIS measurements[16]. A series of specific measurements as part of the Full Function Tests (FFT) provides a reference benchmark for the ICU signals. These tests were performed during the on-ground calibration of MAJIS at the two thermal configurations (126 K,137 K). As stated before, the ICU measurements are also used as an absolute reference for monitoring the in-flight performances. The in-flight tests can be thus compared to the on-ground calibration to monitor any potential change between pre-flight and after-launch over the entire FOV (Section V).

### c. Solid Samples

The other data set of interest for the spectral calibration was acquired with OP5 illuminating different solid samples with a QTH lamp. The samples were spectrally characterized before and after the calibration campaign using a a PerkinElmer FTIR spectrometer. Hyperspectral images of the samples were acquired by scanning them with the hexapod along the direction orthogonal to the MAJIS slit or by commanding the internal scan mirror. Samples' spectra were then extracted and compared to the FTIR spectra (Section III.A.C.3). A companion paper details these measurements and their analyses[17]. Here we focus on the use of this data set for the purposes of the spectral calibration. The spectral calibration analyses rely primarily on two types of images. For the VISNIR channel, acquisitions with the WCS[18] from Labsphere provide the most relevant observation since it exhibits many reference spectral signatures that can be used for the spectral calibration. Most signatures are present at wavelengths < 2.1 µm (see Fig. 5), making the WCS suitable for calibrating the VISNIR channel of MAJIS. Calcite acquisitions were relevant to cover a part of the IR channel. Note that windowing has been routinely applied during these acquisitions.

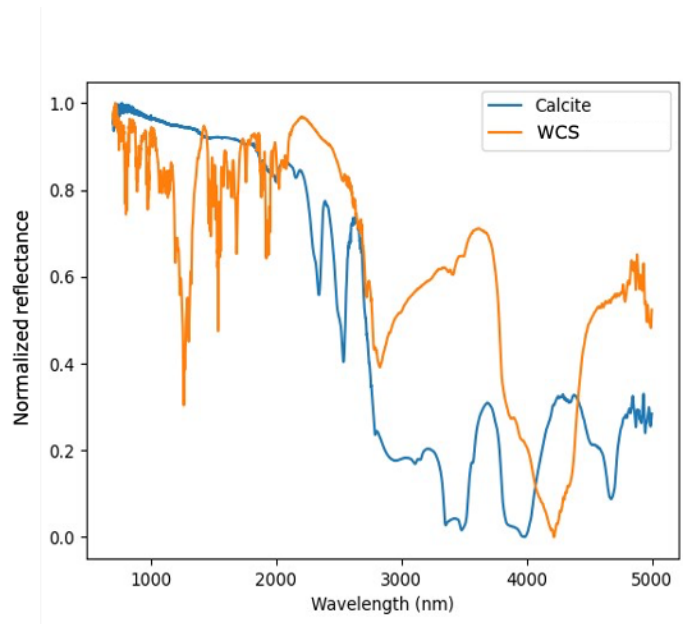

Fig. 5. FTIR spectra of the two solid samples used for the spectral calibration (See text for details).

## III. DATA REDUCTION

### A. High-level data processing

A detailed description of the processing of the MAJIS data will be given in a companion paper[11]. All data are saved as .fits files. Each file is a data cube representing a series of *N* images (2D acquisitions with the same number of rows and columns). The number of images and the size of the images varies from file to file (typically 8 or 16 full-window no process images). In case an acquisition is performed with both channels switched on, each channel is stored in a different .fits file. All files have been processed using the same high-level processing methodology for the calibration analysis.

First, the 3D cube of N images is turned into a 2D median image. This increases the SNR and removes any outliers out of the data set. The proper linearity file, instrument transfer function (ITF) file, and gain are selected based on the acquisition mode and channel[11]. More specifically, the linearity file is applied to correct the H1RG detector's non-linearities[19,20]. Then, the gain and ITF file are applied to transform the measurements reported by each pixel in Analog-to-Digital-Unit (ADU) into a radiance expressed in $(W/m^2/sr/\mu m)$.

### B. Monochromator scans data processing

As stated in Section II.C.1, the monochromator spectral scans provide data across a significant part of the spectral domain of the instrument, namely from 0.7 μm to 3.5 μm. These acquisitions were performed at three or fewer FOV positions (see Table IV and Table V).

1. **Spectral response**

The spectral response describes how the instrument responds to a monochromatic flux of a given wavelength. The key parameter quantifying the spectral response is the FWHM, also called spectral width. As for previous hyperspectral imaging instruments[6], the characterization of the spectral response of the instrument is based on a least-squares fitting using a Gaussian function:

$$G(\lambda) = ae^{-\frac{1}{2}\left(\frac{\lambda-\lambda_0}{\sigma}\right)^2} \quad (1)$$

where $a$ is the height of the curve's peak, $\lambda_0$ is the wavelength corresponding to the center of the peak used for the absolute spectral calibration, and $\sigma$ is the standard deviation characterizing the curve's width used to assess the spectral FWHM. The FWHM is linked to the $\sigma$ parameter (Eq. 1) by the equation below:

$$\sigma = \sqrt{\frac{FWHM}{2\ln 2}} \quad (2)$$

To enhance the SNR and mitigate the effects of inoperable detector pixels, the spectrum data points were constructed by taking the median value of 40 physical pixels in the spatial direction around the target position. Background subtraction was also applied to the data to eliminate influences from the instrument and calibration setup that are uncorrelated with the monochromatic source. This correction involved subtracting a frame captured under identical conditions (integration time, operative modes), but with the light source switched off. In instances where background data were unavailable, background information obtained from another monochromator scan was utilized. Subsequently, a Least-Square (LSQ) algorithm was employed to fit this background-subtracted signal to a Gaussian function, enabling the extraction of the FWHM.

Fig. 6 shows examples of comparison between scan profiles and their best fits. Some discrepancies can be observed, especially near the maximum of the response. However, the difference between the fitted FWHM and the data FWHM is negligible in most cases, meaning that a Gaussian function is appropriate to represent the spectral response function. Using such a function provides a more accurate estimate of the actual width if a few spurious values are present in the data. We estimate that this algorithm provides the FWHM with an accuracy better than 0.2 nm.

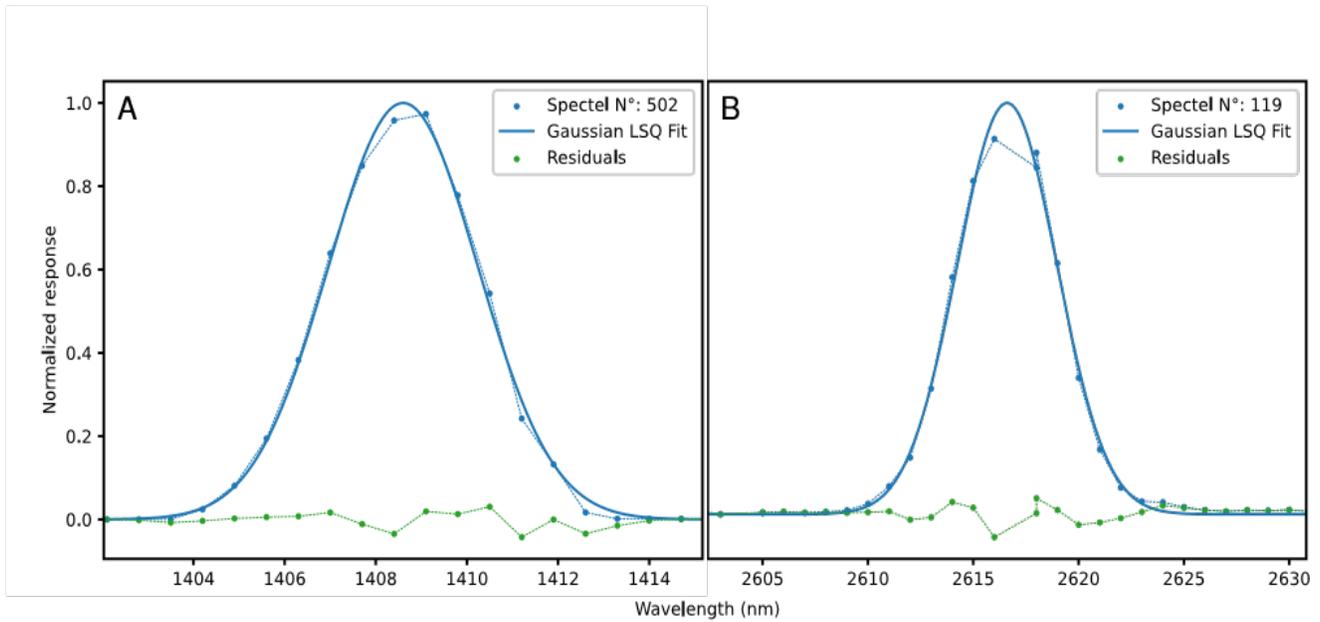

Fig. 6. Example of a Gaussian fit of the OP1 data for the response of a physical spectel at the FOV position 200 to successive monochromatic measurements. (A) VISNIR physical spectel #502 and (B) IR physical spectel #119 (Due to a software error during the calibration campaign, the 2617 nm monochromatic measurement was skipped, and the 2618 nm measurement was repeated).

The number of spectels for which their spectral response is retrieved depends on the spectral domain covered by the monochromatic scan, the step used for the scan, the spectral sampling of the instrument, and the operability of the spectels being tested. Fig. 7 shows an example of the Gaussian fits of a cluster of successive boresight spectels in the 1400-1425 nm range.

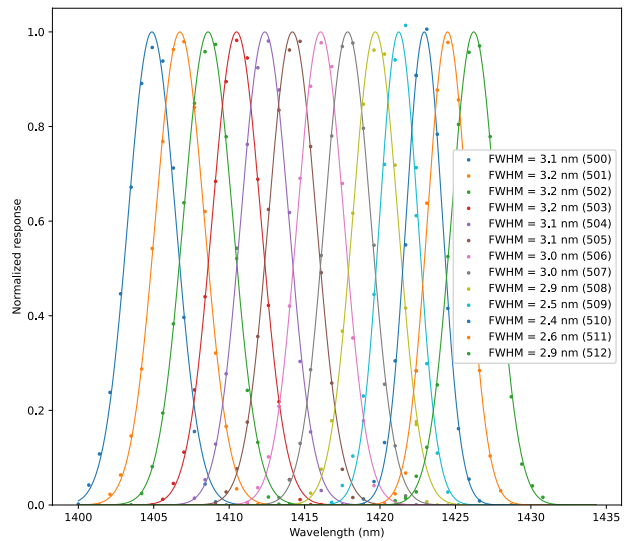

Fig. 7. Gaussian fits of physical spectels response to a monochromatic scan of the 1400-1435 nm range at the FOV position row 200 (boresight) performed during the calibration campaign.

### 2. Absolute spectral calibration

In addition to the spectral response characterization, a second parameter of interest derived from the analysis of the monochromator scans is the Central WaveLength (CWL). It corresponds to the spectral position of the maximum of the Gaussian, i.e., the parameter $\lambda_0$ of Eq. 1. The fitting algorithm gives low error margins on the determination of $\lambda_0$ (< 0.1 nm), one order of magnitude below the precision of the monochromators. Each derived value of CWL provides a data point for the absolute spectral calibration based on other measurements since monochromator scan measurements are restricted to a few wavelengths and FOV positions.

## C. Spectral data processing

### 1. Atmospheric data processing

The measurements acquired with the OP3 setup consist of background-subtracted spectra. The telluric signatures were then fitted to the HITRAN[15] model (considering a temperature of 296 K and a pressure of 1 atmosphere) to provide their absolute spectral positions. This approach provides a very accurate spectral source independent of the setup spectral performances. This leads to high accuracy on the absolute calibration data for strong signatures (< 0.5 nm).

More specifically, the methodology for calibrating the spectral central wavelengths of the instrument consisted of the following steps. A background spectrum was acquired with no gas in the optical path and subtracted from the raw data. The HITRAN[15] model data were convoluted to a Gaussian window with an FWHM of 3.5 nm for the VISNIR channel and 6.0 nm for the IR channel to match the average instrument spectral FWHM. Wavelength ranges exhibiting strong absorption features were selected based on the convoluted HITRAN data. The continuum was subtracted from the data on these intervals, allowing a match between the instrument data and the HITRAN data to determine the spectral central wavelengths.

The matching procedure is automatically performed using the algorithm presented in the following. For the selected wavelength range and spectels, which correspond to one or more absorption signatures, two parameters are optimized: the central wavelength of the first spectel and the sampling between the relevant spectels (i.e., the constant wavelength step between successive spectels). The objective (or cost) function of the optimization algorithm associates a wavelength to each pixel using these two parameters and interpolates the model transmittance at these wavelengths. It then calculates the Sum of Squared Differences (SSD) between the measured and interpolated transmittance. The optimization goal is to find values for the optimization parameters that minimize the SSD. The initial values and boundary settings of the algorithm have been selected based on theoretical models and preliminary analyses. The solver uses a Broyden–Fletcher–Goldfarb–Shanno algorithm to minimize the cost function[21]. This processing is expected to find values for the optimization parameters that provide a good match between the model and measured datasets while ensuring that successive spectels are separated by a constant wavelength step, which is a reasonable approximation over spectral ranges of a few tens of μm (the relevance of this approximation is checked afterward with the actual spectral variations of this wavelength step, see Fig. 13). However, in practice, this algorithm is often applied to a small number of signatures to simplify the problem and avoid any overfitting artifacts (Fig. 8 (A)). The quantification of errors on the spectral positions is not straightforward due to a large number of factors such as the noise and uncertainty in the measured data, the accuracy of the model data, the assumptions made by the optimization algorithm, the quality of the cost function, and the accuracy of the initial guess and boundaries of the optimization procedure. Nevertheless, we assess these errors by using a bootstrapping approach. Bootstrapping is a statistical method that generates resampled data to evaluate the error margin. Multiple bootstrap samples are first randomly generated from the original dataset. The optimization algorithm described previously is then run on each of these resampled datasets, and the distribution of results can be used to estimate the variability or uncertainty statistically. To estimate the error margin on the original result, the standard deviation of the resampled distributions is calculated for each spectel (Fig. 8).

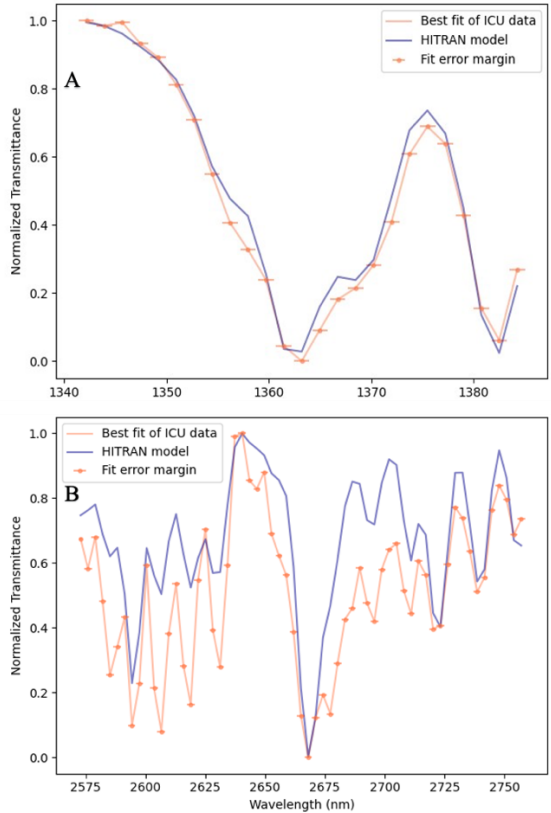

Fig. 8. Panel A: Example of a best-fit match of the OP3 data with the HITRAN-generated transmission spectra near the 1.365 μm water vapor signatures. See text for the explanation on how the error bar on the spectral position is derived for each spectel. Panel B: Same as Panel A but for the IR channel and the 2.65 μm water vapor signatures.

2. ICU data processing

*a. Absolute spectral calibration*

The ICU's didymium and polystyrene filters (see Section II.A.2) introduce several diagnostic absorption signatures on the VISNIR and IR source signals. The sequence of acquisitions with the ICU includes a series of dark acquisitions, background acquisitions, and source acquisitions. The same algorithm was used as the one used to match atmospheric signatures (See Section III.C.1), but in this case, the inputs are the ICU data and laboratory spectra of the filter samples. Thanks to the broad width of the signatures and the excellent SNR, more extensive wavelength ranges could successfully be matched by the optimization algorithm while maintaining low error margins and without falling into a local minimum. One example of the best-fit procedure for each channel is shown in Fig. 9.

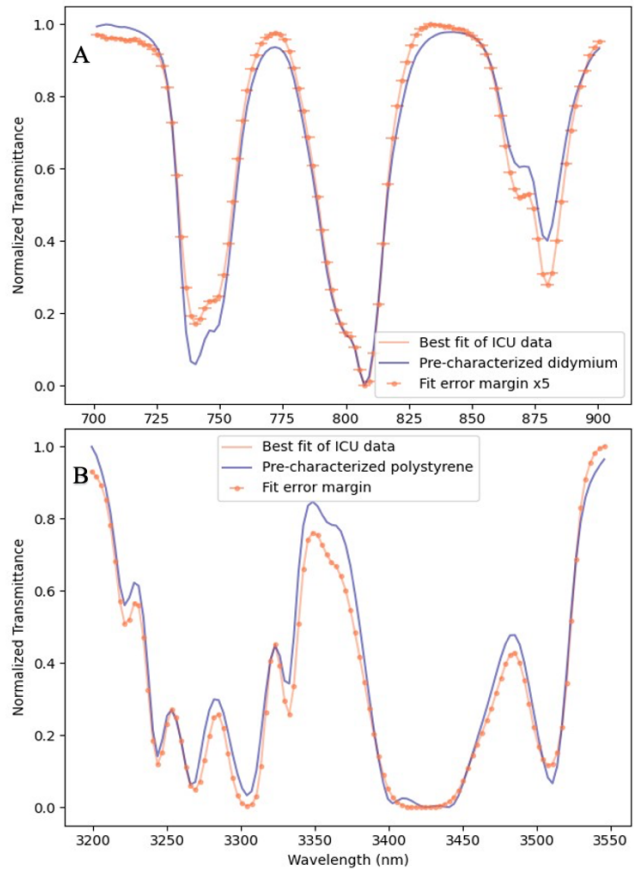

Fig. 9. Panel A: Example of a match between the MAJIS VISNIR ICU data and the reference laboratory signatures of the didymium sample (blue). Panel B: Same as Panel A but for the IR channel and a set of polystyrene signatures near 3.35 μm.

### b. Spectral response

Data from the ICU can offer insights into the relative changes in the FWHM of the response. This is done by convoluting the reference characterization data (from didymium and polystyrene) with Gaussian functions of different FWHM values. Once a specific signature is chosen, its shape is compared to the data gathered by the ICU. The quality of the match between these data sets is assessed by calculating the SSD between the ICU data and the convoluted data. The FWHM of the instrument is then determined to be the value that minimizes this SSD. However, this approach is limited by the broad shape of the signatures to give an accurate absolute estimate of the FWHM. Moreover, eliminating the continuum can introduce a slight bias in absolute terms. Nevertheless, by processing the ICU data in the same way in every case, any relative changes in FWHM can be assessed. In other words, while this method may not be highly accurate, its level of precision is sufficient to yield valuable insights into the evolution of the FWHM.

### 3. Solid sample data processing

The same methodology as the one described for the atmospheric gas acquisitions (Section II.C.1) was applied to these images to extract the absolute spectral positions before deriving the performances of the instrument. Backgrounds acquired for the sample measurements are subtracted from the sample acquisitions. The WCS fitting results show that several signatures can be matched across a large part of the spectral domain (from 1100 nm to 2000 nm) with low error margins (< 0.5 nm). The calcite sample results provide relevant absolute spectral positions restricted to the lower wavelengths (< 3000 nm) of the IR channel (Fig. 10). This is due to the low signal emitted by the light source on OP5 compared to the thermal emission of the samples themselves (samples at ambient temperature, with high emissivity), which precludes the identifications of the spectral signatures at longer wavelengths.

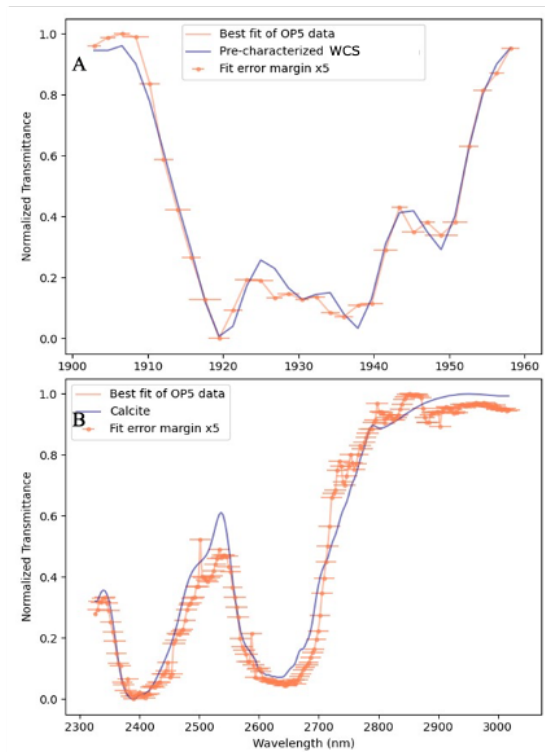

Fig. 10. Panel A: Example of the best match procedure with the WCS sample (orange) around 1.93 μm signatures. The reference absorption of the sample is shown in blue. Panel B: Same as Panel A but for the calcite sample in the IR channel.

## IV. RESULTS

The results presented in this section correspond to the nominal thermal case (130 K at LDO premises and 126 K at IAS, see Section II.A.3.)

## A. Absolute spectral calibration

The absolute spectral calibration is characterized by two metrics (Table I) at the physical spectel level. The baseline of the analysis (called option 1 hereafter) is to gather all measurement points for each channel (Table IV, Table V and Table VI) to fit an absolute calibration model that minimizes the residuals. The instrument's design leads to a quasi-linear spectral calibration. Hence, to account for the deviations to linearity of the grating dispersion law, a polynomial function becomes the choice for fitting the calibration data. Specifically, a fourth-degree polynomial function was chosen as it accurately fits the data points across both channels while limiting overfitting artifacts. Second-degree polynomial fitting results in a significantly larger Sum of Squared Errors (SSE) (multiple data points are consistently 1 nm or further away from the fitting polynomial). Higher-degree polynomial fitting improves the SSE only marginally at the expense of much more significant overfitting artifacts at the edges of the wavelength range. Third-degree polynomial fitting also provides a good balance with a relatively low SSE without overfitting artifacts, but the fourth-degree polynomial was preferred for its overall slightly better consistency with the data and theoretical sampling.

We noticed that the data points acquired with OP1-LDO and OP5 exhibited a distinct trend compared to those for OP0, OP1-IAS, and OP3 for both channels. In particular, the best fit at wavelengths greater than 1400 nm misses many data points of these sources even when accounting for their error margins. Therefore, we considered a second option in which the minimization procedure was performed using only the data points from OP0, OP1-IAS, and OP3. The restricted data set of this option (called option 2 below) was a priori given higher confidence for several reasons. The atmospheric signatures can be taken with increased confidence, and the ICU (OP0) is specifically designed for the spectral calibration of the instrument. For the IR channel, we also investigated an option 2 without OP1-IAS data because the data were inconsistent with atmospheric signatures by several nm.

For both channels, the solution provided by option 1 has high uncertainty, as expected, while the fit with option 2 matches very well with all the data points, suggesting that these data sets are more consistent. This leads to low error margins in the resulting calibration (~ 0.7 nm for the VISNIR channel and ~ 2 nm for the IR channel). The results of the derived CWL are shown in Fig. 11 and Fig. 12 for the center of each channel. The polynomial coefficients ($a\_0$ being the coefficient of the constant term and $a\_4$ the coefficient of the fourth-degree term) as well as the resulting spectral ranges are summarized in Table VII.

Table VII. Spectral range comparison for both channels and both options

|  |  | **Spectral Range (nm)** | **Error Margin (nm)** | **a_0** | **a_1** | **a_2** | **a_3** | **a_4** |
|---|---|---|---|---|---|---|---|---|
| **Option 1** | VISNIR | 489 – 2349 | 2 | 4.886e2 | 1.795 | 2.310e-4 | -3.322e-7 | 1.401e-10 |
|  | IR | 2268 – 5556 | 3.5 | 2.269e3 | 2.991 | 4.093e-4 | -3.138e-7 | 1.502e-10 |
| **Option 2** | VISNIR | 491 – 2358 | 0.7 | 4.902e2 | 1.768 | 3.639e-4 | -5.518e-7 | 2.604e-10 |

| | IR | 2271 – 5555 | 2 | 2.270e3 | 2.991 | 3.801e-4 | -2.536e-7 | 1.170e-10 |

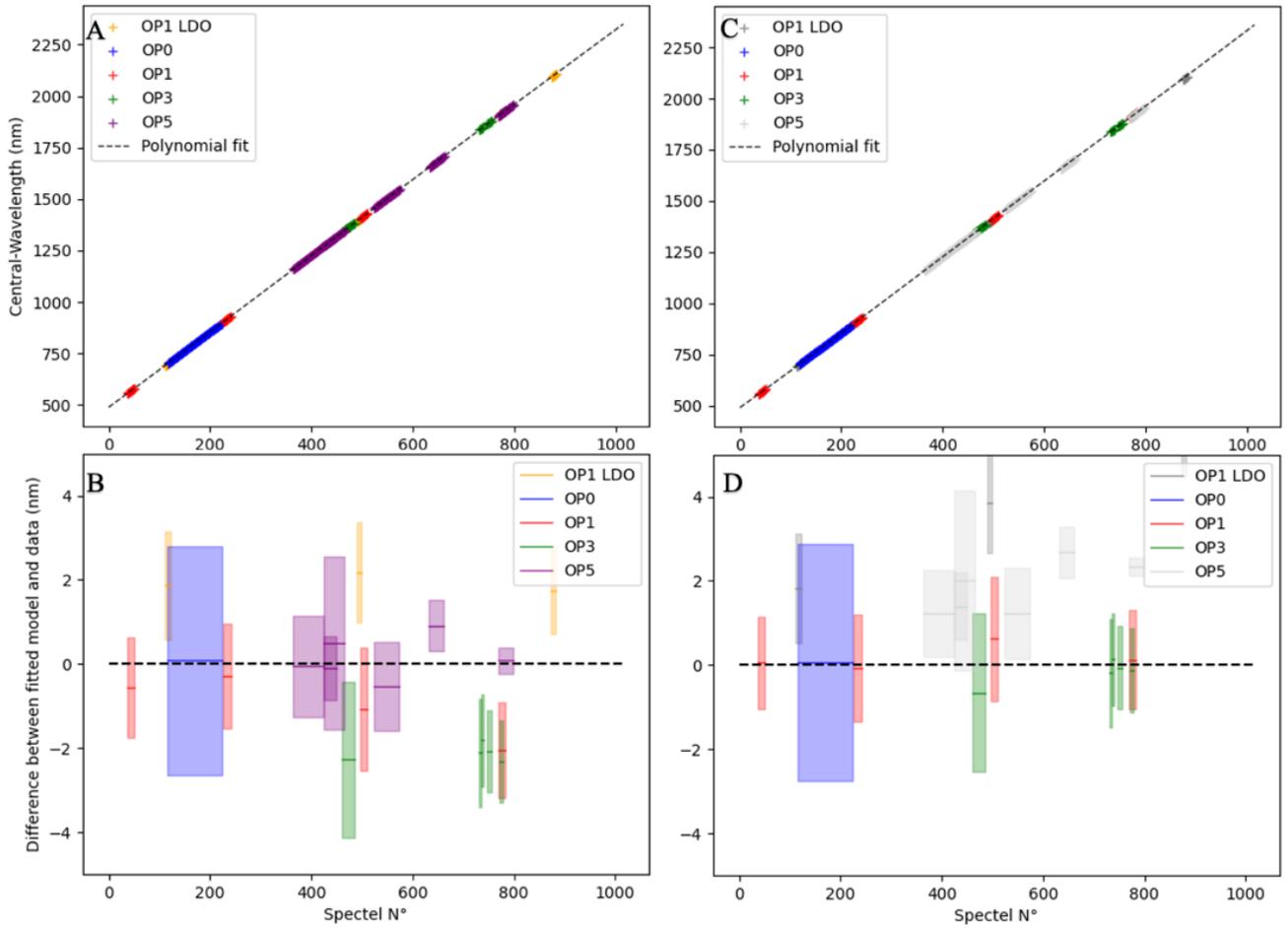

Fig. 11. VISNIR absolute spectral calibration (left panels: option 1; righ panels: option 2) for the FOV position 200 at the nominal temperature. These panels compare spectel numbers to their associated CWL. (A) Fourth-degree polynomial fit of the central positions of the spectels for the VISNIR channel with all calibration data considered ("option 1", see text for details). (B) Comparison between the derived CWL and the measurements. One box corresponds to one series of measurements (i.e., multiple data points). The box size in the X axis indicates the spectral range of the measurement series, and the box size in the Y direction shows the error in the spectral position of the data points. The position of the middle line represents the fitted model considered. As the wavelength increases, so do the spectel numbers. (C) and (D) A & B respectively, but for a selection of calibration data only ("option 2"; unused calibration data are in light grey).

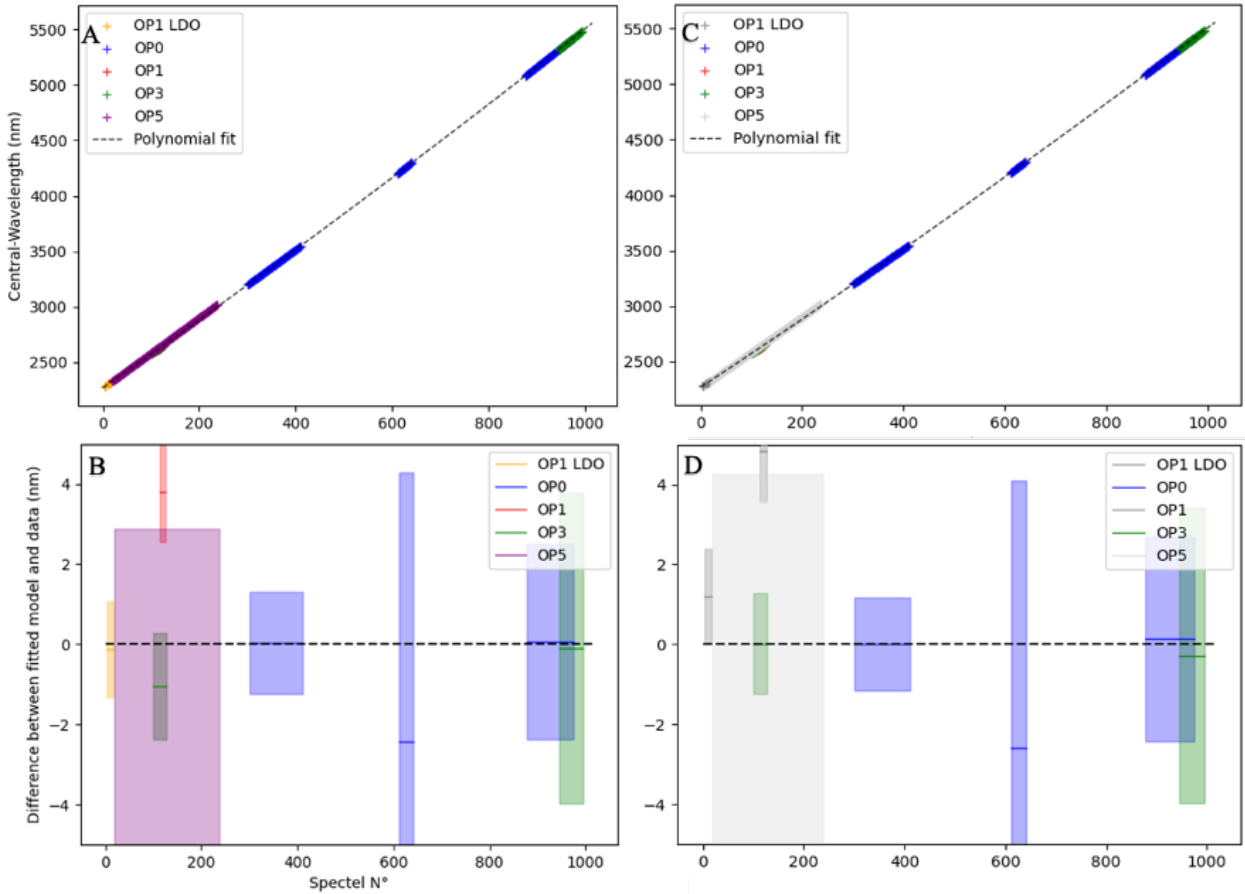

Fig. 12. IR absolute spectral calibration (both options) for the FOV position 200 at the nominal temperature. All panels (A, B, C, and D) are the same as Fig. 11 but for the IR channel.

The resulting spectral samplings of both channels (i.e., the difference in CWL between two successive spectels) are well in-line with the pre-calibration OH models (Fig. 13). For the VISNIR channel, option 1 has the closest overall match. The fitting only deviates from the model at the edges of the spectral domain (within a 2% deviation for the rest). Option 2 is an excellent match for lower wavelengths (<1750 nm) but differs significantly at higher wavelengths. It should be noted that the last data point for this option is around 1950 nm; therefore, the deviations observed at higher wavelengths can be attributed not only to the data but also, to some extent, to the choice of a fourth-degree polynomial fitting. The third-degree alternative resulted in an opposite trend in the sampling, leading to significant differences at the edges (up to 3 physical spectels of difference on the absolute calibration at the end of the wavelength range). The third-degree alternative has the drawback of further deviating from the OP1-LDO data points (discarded for this option), whereas the fourth-degree alternative provides satisfactory fits of all data. For the IR channel, the sampling is virtually identical for both options, remaining within a 2% deviation from the pre-calibration model. Note that the third-degree polynomial fitting gives results virtually identical to the current calibration (<1 physical spectral of variation across the whole spectral range).

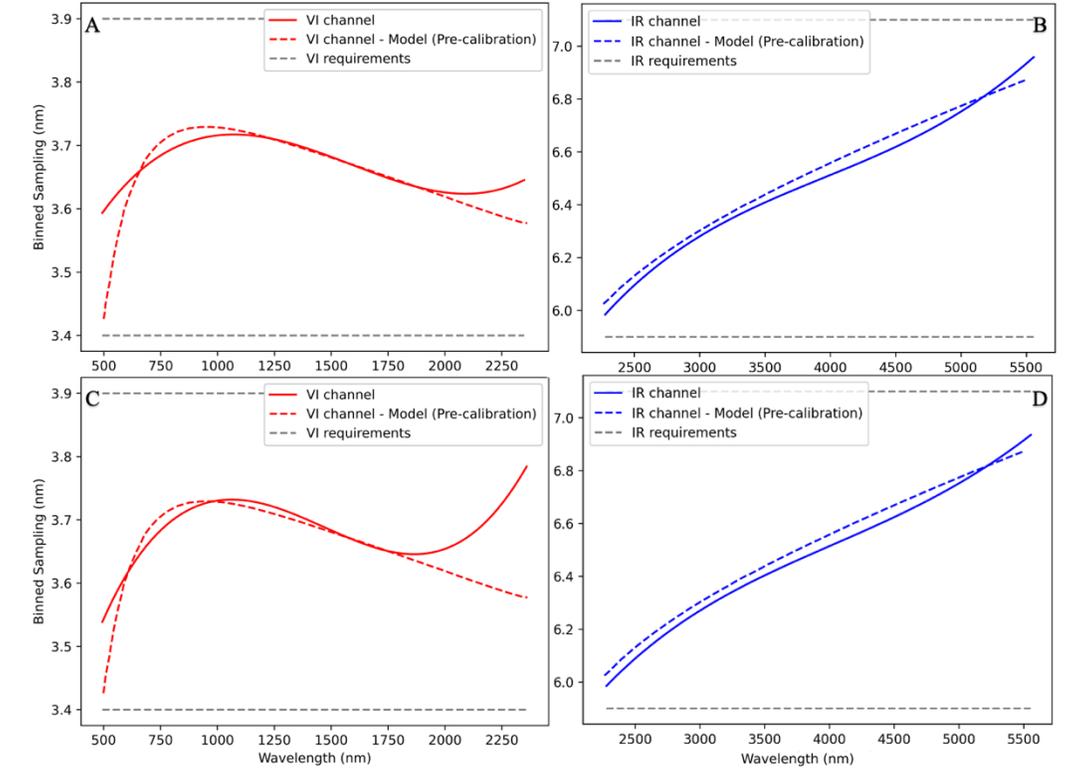

Fig. 13. Spectral sampling of nominal spectels for the two considered dataset options at nominal temperature and in the center of the FOV of both channels. (A) Spectral sampling of the VISNIR channel based on the polynomial fitting results of the absolute calibration provided by option 1. The grey dashed lines indicate the requirements, and the colored dashed line shows the sampling expected based on the theoretical model of the OH performances developed during the design phase. (B) Same as (A) but for the IR channel. (C) Same as (A) but for option 2. (D) Same as (C) but for the IR channel.

## B. Spectral response at the physical pixel level

As mentioned in Table IV and Table V, three positions along-slit are covered (FOV position row 20, 200, 380) by six monochromator scans (seven for the boresight position) for the VISNIR channel. The same three positions along-slit for the IR channel are covered, but only two scans below 2700 nm are available due to the significant thermal emissions at higher wavelengths. As distinct trends are observed between LDO and IAS data sets, we report the FWHM separately derived from each data set (Fig. 14).

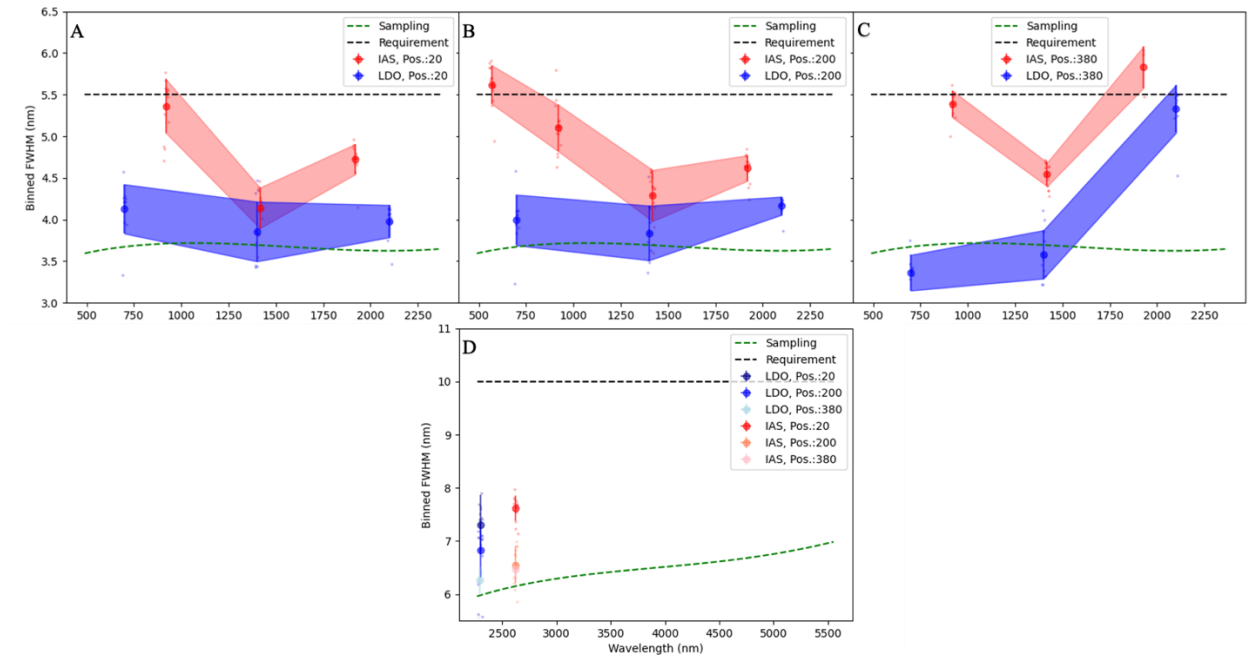

Fig. 14. LDO- and IAS-based FWHM vs wavelength. (A) Evolution of the FWHM across the spectral range of the VISNIR channel for the data acquired at LDO and IAS for FOV position 20. The sampling (green dashed line) is shown to highlight any potential correlation with the FWHM. (B) Same as (A) but for position 200. (C) Same as (A) but for position 380. (D) Same as (A) but with all FOV positions of the IR channel.

For the VISNIR channel, the FWHM measured at LDO is well within requirements but is significantly lower (10 to 40%) than the FWHM measured at IAS. The difference is more significant for low (<1000 nm) and high (>1700 nm) wavelengths. We hypothesize that this discrepancy is primarily related to the fact that the entire entrance baffle of MAJIS was illuminated at IAS while this was not the case at LDO. More straylight was observed with the entirely illuminated entrance, which is consistent with the wider spectral dispersion observed with the IAS setup.. The FWHM difference between IAS and LDO appears to increase as we move away from the center columns and the middle of the FOV. This difference is maximum at wavelengths <1000 nm. This is consistent with the straylight observed on MAJIS (less straylight in the FOV center than on its edges and straylight maximum at <1000 nm [11]). As MAJIS entrance baffle will also be fully illuminated at Jupiter, this suggests that the actual performances of MAJIS will be more similar to those measured at IAS than those measured at LDO. The IAS-based FWHM is slightly out of specification at short wavelengths. We have also explored the possibility that IAS OP1 monochromator FWHM may have been wider than expected (up to 2 nm instead of 0.7 nm for VISNIR wavelengths, as discussed in a companion paper[5]). The lowest difference between IAS and LDO is observed at 1400 nm for position 20 (Fig. 14.A): the difference between LDO (FWHM 3.85 nm) and IAS (FWHM 4.15 nm) requires a monochromator FWHM of 1.55 nm, which could be consistent with this putative wider IAS OP1 FWHM. However, in most cases, differences would require that the IAS OP1 monochromator FWHM was well above 2 nm, in particular for wavelengths <1000 nm, where the contribution of the straylight remains the most plausible explanation. Finally, we note that there is no obvious correlation between the sampling and the FWHM of the VISNIR channel.

For the IR channel, fewer measurements are available (only 2 per FOV position). However, no significant difference between IAS and LDO data is observed, and the measurements are compliant with the requirements. No definitive conclusion can be made for higher wavelengths due to the lack of measurements. Still, given the low FWHM variation across 300 nm, the substantial margin between the performances and the requirements at low wavelength, and the absence of stray light for the IR channel, it is anticipated that the FWHM will meet the requirements across the entire IR spectral range.

### C. Effect of the binning on the spectral response

The response of the physical spectel is modeled using a Gaussian function. Binning the data of 2 physical spectels shows that a Gaussian function also provides a satisfactory fit of the instrument's spectral response for nominal spectels. Fig. 15 compares the response of a physical (oversampled) spectel and a nominal spectel (2 physical spectels binned). The FWHM is multiplied by a factor of approximately 1.215 compared to the physical (oversampled) spectel. A large portion of the calibration spectral data (~ 100 spectels) was tested. The measured average factor increase of the FWHM was $1.21 \pm 002$. This factor is in line with the expected theoretical factor (factor of 1.21) resulting from the numerical summation of two Gaussian functions, which have a central wavelength, $\lambda_0$, separated by the theoretical spectral sampling of the instrument. The $\lambda_0$ parameter for the nominal spectel is found to be simply equal to the averaging of the corresponding parameters in the two Gaussian responses binned.

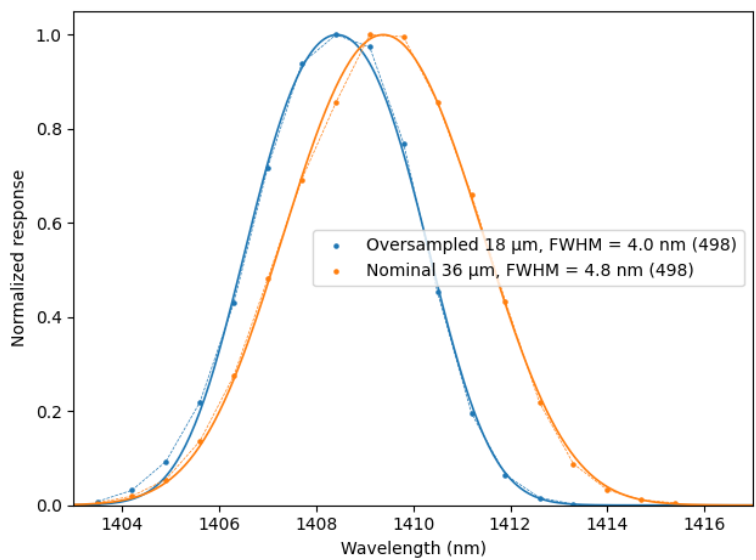

Fig. 15. Comparison of the oversampled and nominal spectral response of the VISNIR channel around 1.4 µm (physical spectel N°498 (out of 1016) in blue and its associated nominal spectel in orange). The retrieved FWHM for the nominal case is very close to the value of the FWHM of the nearby physical pixel multiplied by the theoretical factor of 1.21.

On the other hand, the response can no longer be modeled using a Gaussian function for larger binning modes of the nominal spectel ×2 (leading to a 72-μm pixel size) and ×4 (leading to a 144-μm pixel size). Fig. 16 shows the effect of larger spectral binning modes on the shape of the response. All derived values of FWHM in Fig. 16 are close (within 5%) to the theoretical values that could be calculated from the proportionality coefficients obtained from numerical summations of Gaussian functions and listed in Table VIII.

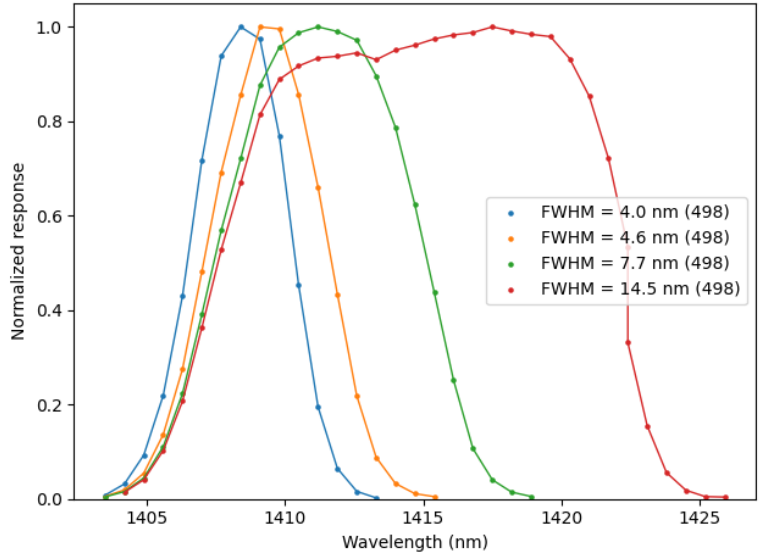

Fig. 16. Examples of the spectral responses for the four spectral binning modes (oversampling (blue), nominal (orange), binning ×2 (green), and binning ×4 (red)). Column 498 (in terms of the first physical pixel position) corresponds to 1408.5 nm.

In the case of binning ×2 and ×4, the proposed modeling approach uses a function defined as the absolute difference between two Cumulative Distribution Functions (CDFs) of the normal distribution, normalized by its maximum value. This Gate-Gaussian function is characterized by its center, the Gaussian component's standard deviation, and the Gate function's width. As for the nominal binning, the FWHM for the ×2 and the ×4 binning modes can be estimated from the spectral response of the physical spectel by applying a factor of 2.03 and 3.60, respectively (Table VIII). Fig. 17 shows this methodology applied to the spectral binning modes ×2 and ×4, corresponding to the summation of 4 and 8 physical spectels, respectively. The function used for the fitting matches the experimental data well in both binning cases, leading to a FWHM from the experimental data virtually identical to the FWHM of the fitted function. The standard deviation around the expected theoretical value resulting from the modeling of about 100 spectels is ~5%.

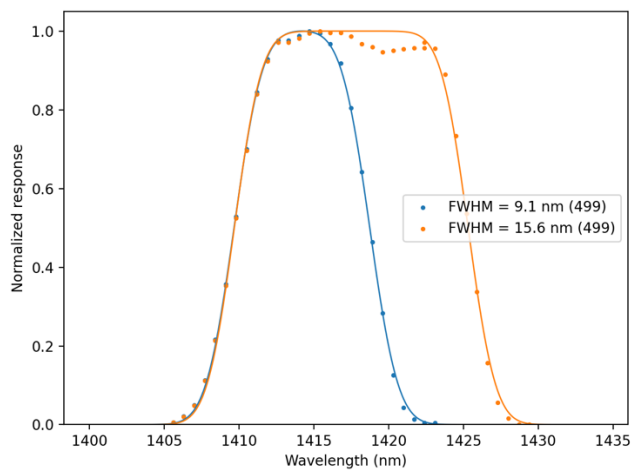

Fig. 17. Fitting of the spectral response in the case of binning mode ×2 (blue plot) and ×4 (orange plot) at 1420 nm. A Gate-Gaussian function (representing a gate function with Gaussian edges) reproduces the measured spectral response in these cases. The FWHM ratio (~1.7) is consistent with the theoretical proportionality parameter (1.77) from Table VIII.

Table VIII. Spectral response models and FWHM proportionality parameters for the MAJIS spectral binning.

| Binning Mode | Function | Proportionality |
|---|---|---|
| Oversampling | Gaussian | 1 |
| Nominal | Gaussian | 1.21 |
| ×2 | Gate Gaussian | 2.03 |
| ×4 | Gate Gaussian | 3.60 |

### D. Smile

The smile effect is caused by optical aberration and results in a variable curvature of the slit image across the spectral range[6]. It can be estimated by comparing, for a given column, the CWL of any pixel along this column w.r.t. to the CWL measured on the middle of the FOV at R=400, which can be expressed as:

$$\text{Smile shift correction}(R, C) = CWL(R, C) - CWL(400, C) \quad (3)$$

$R$ and $C$ indicate the position of a given physical pixel in the FOV in terms of row (spatial index range from 1 to 800) and column (spectral index range from 1 to 1016), respectively (see Fig. 1). Some spectral signatures used for the absolute calibration could not be used for the smile correction due to the low magnitude of the smile corrections (< 1 nominal spectel) and the low SNR precluding an accurate characterization along the FOV. The VISNIR channel smile was characterized using monochromator scan, atmospheric, sample, and ICU data (Table III). Because of the high ambient thermal background occurring during the ground measurements, the IR channel smile was described using atmospheric signatures near 2660 nm only. Still, thanks to the polystyrene signatures in the ICU acquisitions, the smile analysis was extended at around 3370 nm. For all of the measurements except for the ICU and atmospheric measurements, three FOV positions were available.

The methodology used to derive the CWL along the FOV is the same as the one used for deriving the absolute spectral calibration (IV.A). The estimation of the spectral shift using the monochromatic measurements for the VISNIR channel is summarized in Table IX. It must be noted that the lower accuracy of the monochromator measurements (about 1 nm) makes it challenging to assess more minor magnitude distortions reliably.

Table IX. Monochromator scan scans smile table summary. The values indicate the shift in nm w.r.t. the center of the FOV according to Eq (3) for three wavelength ranges

| Smile (nm)   | 700-900 nm      | 1400 nm      | 1900-2100 nm  |
|---|---|---|---|
| FOV Row 20   | $1.5 \pm 1.00$ nm  | $1 \pm 1.00$ nm | $-1 \pm 1.00$ nm |
| FOV Row 380  | $-0.5 \pm 1.00$ nm | $1 \pm 1.00$ nm | $2 \pm 1.00$ nm  |

In the series of acquisitions for which the FPA frames are not "windowed" to a certain part of the FOV, the smile can be consistently estimated across the entire FOV. This is achieved using the ICU and certain OP3 acquisitions (Fig. 18). Such data set can be used to establish a 2D map of the smile correction by using a second-degree polynomial interpolation (this provides a smoother correction that is less sensitive to algorithmic noise), first in the spatial direction and, then, in the spectral direction. Due to the limited data from the IR channel, only a 1D correction was derived by using a second-degree polynomial interpolation in the spatial direction and a constant value in the spectral direction based on the average of the two smile corrections provided by the ICU and atmospheric signatures (Fig. 19). The map correction of the VISNIR channel is almost line-symmetrical w.r.t. the diagonal of the FPA. It is maximal at the corners of the FPA and decreases towards the center "cross". For both channels, along-FOV corrections mostly follow linear trends and not second-degree polynomial trends, unlike the "smile" analogy typically indicates.

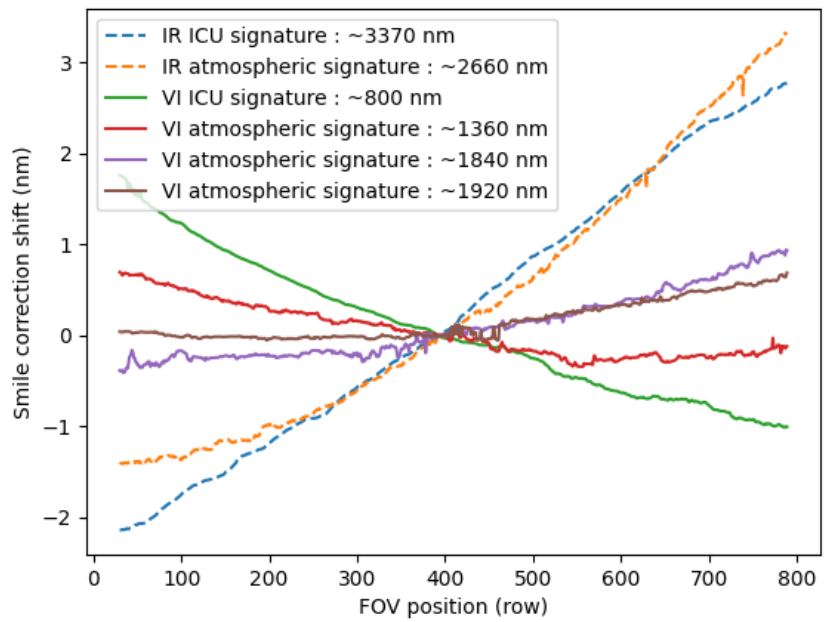

Fig. 18. Smile characterization of both channels for full FOV measurements. The smile characterization uses both channels' ICU and OP3 measurements.

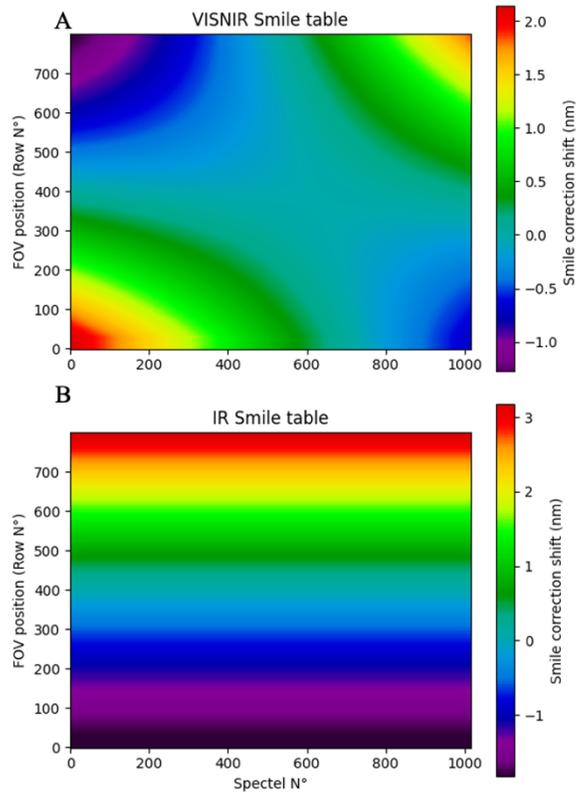

Fig. 19. Mapping of the smile correction factor to be applied for removing spectral smile aberration. A physical pixel scale is considered. Panel A is for VISNIR and Panel B for IR. The shift correction of the smile is to be added to the reference CWL (that correspond to FOV position 200 = 400 physical) to derive the CWL of any physical pixel. The effect of this correction is showcased in Fig. 20.

An example of smile characterization and correction using atmospheric signatures is shown in Fig. 20.

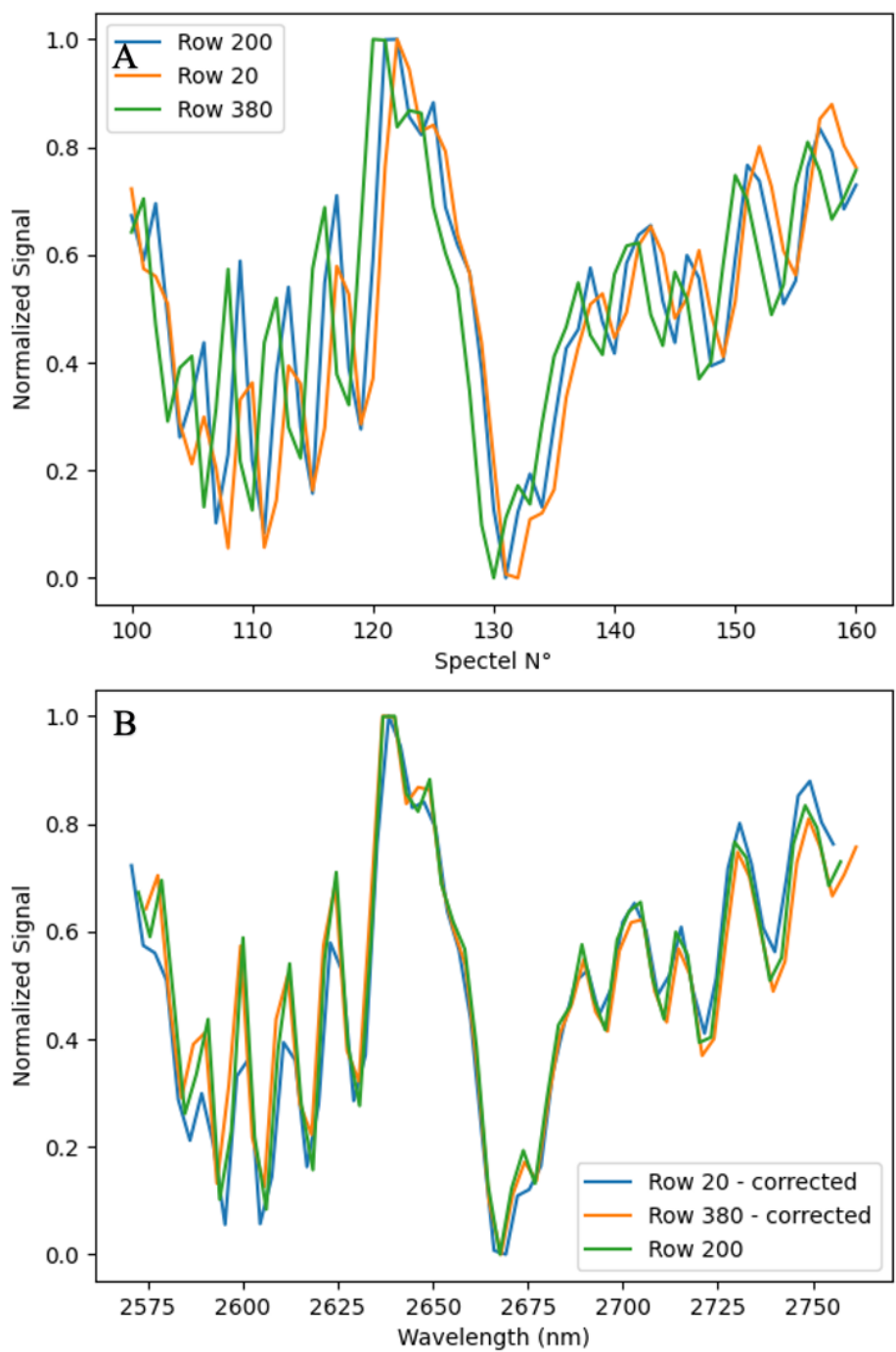

Fig. 20. Illustration of the smile shift correction being applied to an atmospheric (water) signature. (A) Example of the smile effect using atmospheric signatures. The three FOV positions spectra are superposed and show that the signatures are slightly shifted. (B) The three FOV position spectra are superposed after the automatic correction (using shift presented in Fig. 19) of the smile effect has been applied.

### E. Impact of OH temperature

While the MAJIS optical bench, optical elements, and their relative mountings have been designed to maintain an isothermal profile in order to keep optical performances within the requirements, the OH spectral response is expected to manifest small changes with the temperature. Therefore, specific measurements under thermal stress (hot case OH and VISNIR detector at 137 K, IR detector at 97 K) were conducted to explore whether the performances of the instrument remained within the requirements (Section II.A.3).

#### 1. Absolute spectral calibration

The absolute calibration is compared using the data sets acquired with the ICU as well as the OP1 and OP3 setups. Due to planning constraints, the other optical paths were not used in the hot case. The OP1-LDO dataset acquired in the lower and upper operative temperature cases is also used to estimate the relative shift in the central wavelength of the explored spectels across a larger OH thermal range. Each thermal case has a dataset that covers different spectels. For this reason, the trend shown by the comparison between the lower and upper operative temperature cases is used as a relative reference. No significant difference in the absolute calibration of the VISNIR channel between the two thermal configurations is found. The OP1-LDO dataset confirms that no relative change is observed between both cases for the VIS-NIR channel. Conversely, the four data sets show a shift in the absolute calibration of the IR channel. The spectral shift to be applied to the hot case data for correcting the spectral calibration was thus derived (Fig. 21). A constant shift of about 2.0 nm is observed at wavelength < 4200 nm between the two temperatures (126 K and 137 K). For the ICU signature around 5.2 µm (physical spectel N°910) and the atmospheric signature around 5.4 µm (physical spectel N°970), an inversion in the shift is found (Fig. 21). However, associated error bars are very large and cover largely the shift derived at shorter wavelength. Actually, because of the increasing thermal background of the instrument for the hot case, these signatures are poorly resolved, leading to a more significant error in the CWL calculation at higher wavelengths (Fig. 22). These points are thus excluded in the final estimation of the spectral shift. If we assume a linear trend with temperature to interpolate and extrapolate these two measurements, the spectral response is drifting with temperature by + 2.0 nm ± 0.3 nm per 11 K, leading to the relationship $CWL(126\ K + \Delta T) = CWL(126\ K) + \Delta T * 0.18\ nm\ /\ K$). In addition, the OP1-LDO data confirmed that a shift can be observed in the CWLs of the spectels of the IR channel between the lower and upper operative temperature cases (II.A.3). The relative shift computed between both cases is in line with the IAS dataset (the shift is equal to + 8.5 nm ± 0.3 nm per 40 K or + 0.21 nm per K).

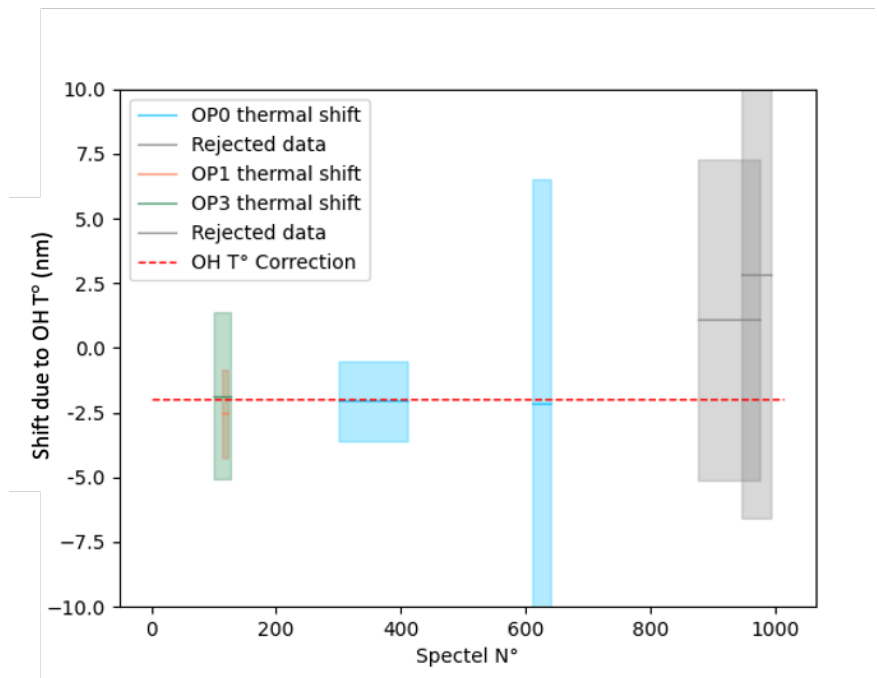

Fig. 21. Shift of the spectral signatures (opposite to the shift of the CWLs) observed during the hot case relative to the nominal case for the IR channel

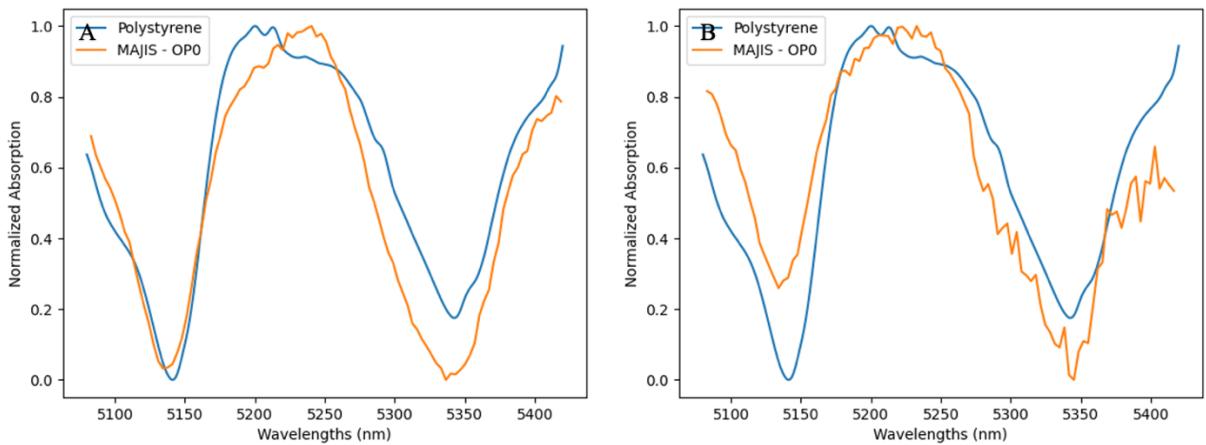

Fig. 22. Illustration of the degradation and the loss in resolution of the polystyrene signature around 5.2 µm between the nominal and hot case resulting from the increasing thermal background. (A) Example of polystyrene signature fitting for the IR channel in the nominal case. (B) Same as (A) but for the hot case. The fitting algorithm performance is significantly less reliable in this case.

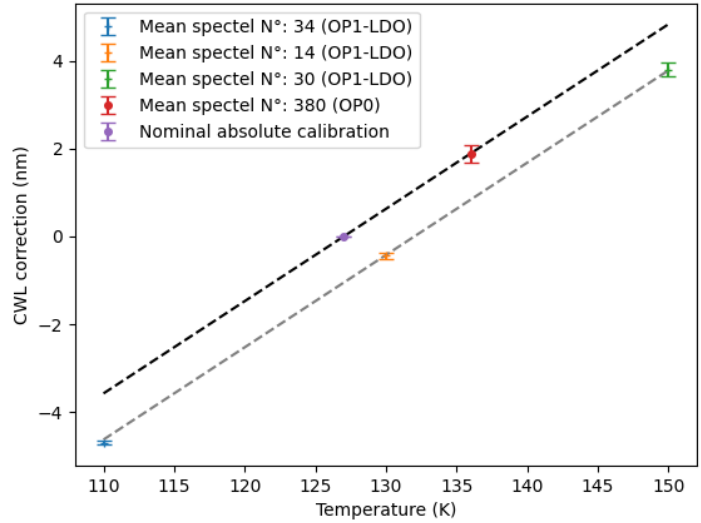

Fig. 23. CWL average shift measured across the tested thermal cases. The shift corresponds to the average difference between the data points at a given temperature and the absolute spectral calibration. Each data point summarizes several measurements. The mean spectel of each dataset is displayed in the legend. It must be noted that the LDO data is used as a relative reference, which is why its spectral position is shifted below the actual calibration (dashed line).

2. **Spectral response**

The FWHM of the spectral response function has been evaluated at three wavelengths for the hot case (OH temperature 137 K) and compared to the nominal case (126 K). An example of a spectel FWHM response is shown in Fig. 24. The results from all available scans are summarized in Table X. For the VISNIR channel, we observe at short visible wavelengths (550 nm) a significant 9% increase of the FWHM with the OH temperature. This increase is reduced to 1% at 1400 nm, a difference that is not significant considering errors. For the IR channel, the increase is 3% at 2600 nm, a value that is again not significant considering error bars. Overall, the analysis thus suggests that the FWHM of the spectral response increases by about 10% from nominal case to hot case at visible wavelengths and may also increase slightly by a few % at other wavelengths. This effect is significant enough to be carefully considered for future in-flight acquisitions.

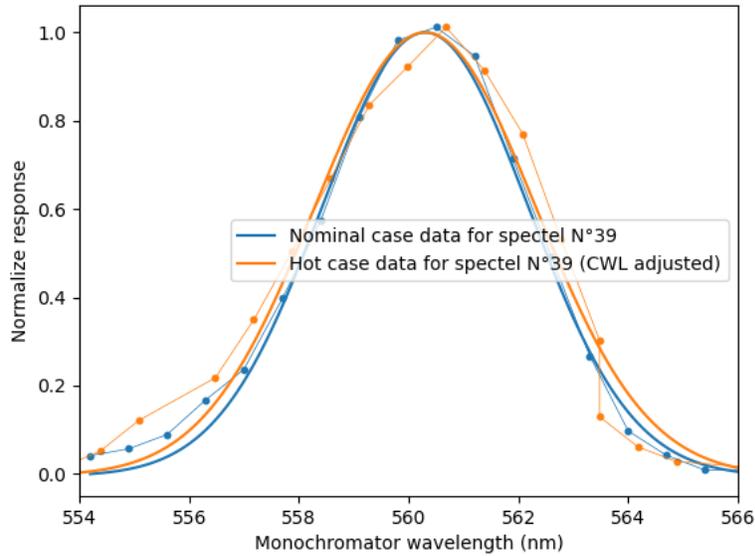

Fig. 24. Example of FWHM degradation due at high OH temperature. The spectral response of the physical spectel N°39 of the VISNIR channel is displayed for the monochromatic scan starting around 565 nm.

Table X. Spectral responses derived for the hot case.

| Scan λ (nm) | OH T° (K) | Mean FWHM (nm) | Std. dev. FWHM (nm) |
|---|---|---|---|
| **550-585** | 126 | 4.64 | 0.19 |
| | 137 | 5.08 | 0.29 |
| **1400-1435** | 126 | 3.54 | 0.25 |
| | 137 | 3.58 | 0.32 |
| **2325-2375** | 110 | 6.72 | 0.05 |
| | 150 | 7.23 | 0.19 |
| **2600-2650** | 126 | 5.41 | 0.29 |
| | 137 | 5.57 | 0.27 |

3. **Smile**

The impact of the temperature on the smile was assessed thanks to the ICU acquisitions. Differences measured (if any) are below the accuracy thresholds of the methodology used. Fig. 25 illustrates the virtually identical smile distortions measured from two ICU signatures. We thus conclude that no correction of the smile effect is required to account for the variation of the OH temperature between the nominal case and the hot case.

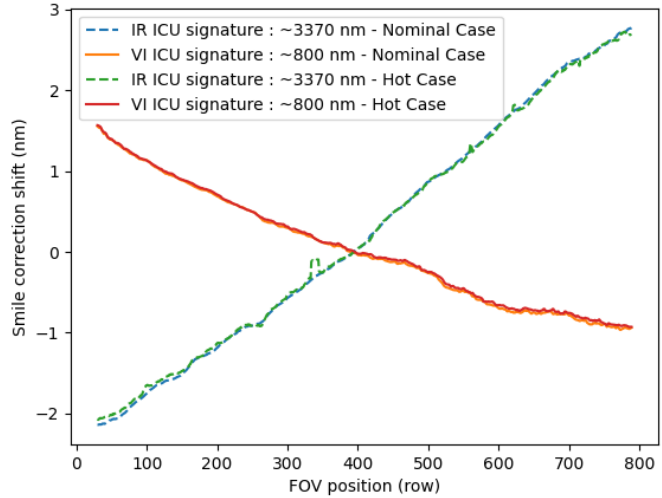

Fig. 25. Comparison between the derived smile correction using the ICU acquisitions for the nominal and hot cases.

## V.   NECP MEASUREMENTS AND COMPARISON WITH GROUND MEASUREMENTS

Measurements obtained during the NECP are discussed in this section. The NECP was the post-launch phase that involved the first operations of all spacecraft instruments, including check-out and calibration activities. During this phase, the OH was at a temperature of 133.5 K. One of the major goals of the NECP is to ensure the integrity of the instruments and the data pipeline via the spacecraft. For MAJIS, the measurements provide some constraints about calibration variation resulting from the various ground tests (thermal, shock, and vibration) performed after the mechanical integration of MAJIS on the spacecraft as well as from the vibrations of the launch and spacecraft separation. These flight measurements are acquired with the ICU subsystem, as no specific pointing was allowed during the NECP. We first compare the pre- and post-launch absolute calibration using the spectral signatures of the didymium (VISNIR) and the polystyrene (IR) filters of the ICU. We observe an absolute calibration shift for both channels (see Fig. 26). This spectral shift is -3.8 nm in the VISNIR and -5.4 nm in the IR; it is constant over the wavelength range available for comparison (between 0.7 μm and 0.9 μm in the VISNIR and 3.2 μm and 3.55 μm in the IR), suggesting it may be constant over each channel. The IR shift is mitigated by the difference in OH T° (see Section IV.E.1), leading to an actual shift of -4.0 nm.

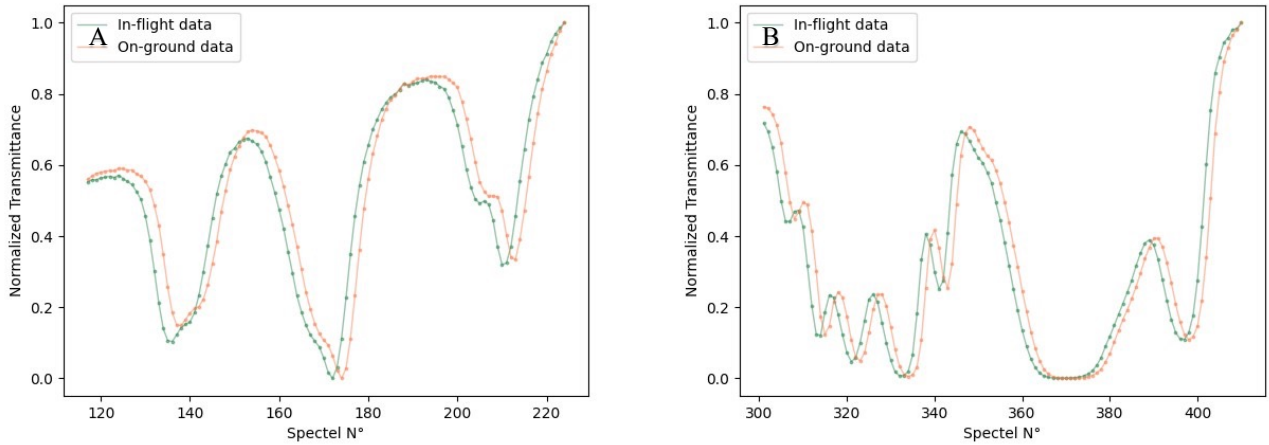

Fig. 26. (A) Comparison of the VISNIR ICU data before (on-ground calibration) and after (NECP) launch. A shift of -2.0 physical spectels (corresponding to a spectral shift of -3.8 nm) is observed. (B) Same as (A) but for the IR channel. A shift of -1.7 physical spectels (equivalent to a spectral shift of -5.4 nm) is observed.

Section III.A.2.b mentions that the spectral response could not be characterized accurately with the NECP data set. However, the spectra measured before and after the launch can be compared using the approach described in Section III.A.2.b. Fig. 27 shows that the spectral response has not changed significantly after the launch (< 10% relative variations), the slight variations being within the uncertainties of the methodology.

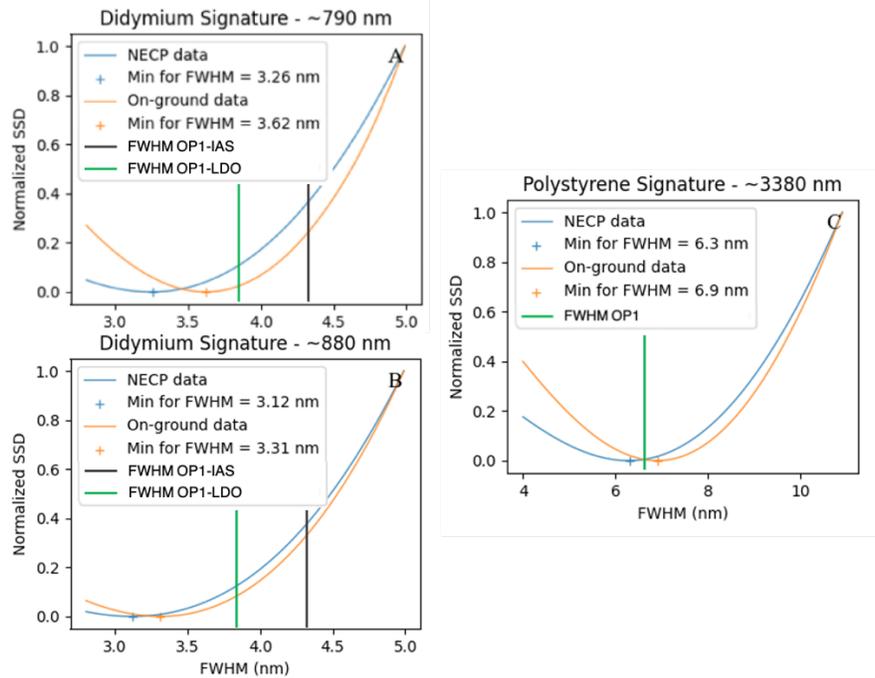

Fig. 27. Examples of the FWHM estimation using an SSD algorithm between on-ground calibration and NECP derived from the convolution of the ICU signatures. (A) Best fits (crosses) for the didymium signature around 790 nm. (B) Same as (A) but for a signature around 880 nm. (C) Same as (A) but for a polystyrene signature around 3380 nm (IR channel).

The algorithm to characterize the smile using ICU data was also applied to the NECP ICU acquisitions. The comparison of the NECP and on-ground ICU data is shown in Fig. 28. Both channels show similar smile trends between the in-flight and on-ground data sets for most of the available data. A small change is observed for the lower parts of the FOV (< Row N°200) of the IR channel with the 3.3 μm signature. At this very edge of the FOV, the smile is reduced by about 22% (0.5 nm).

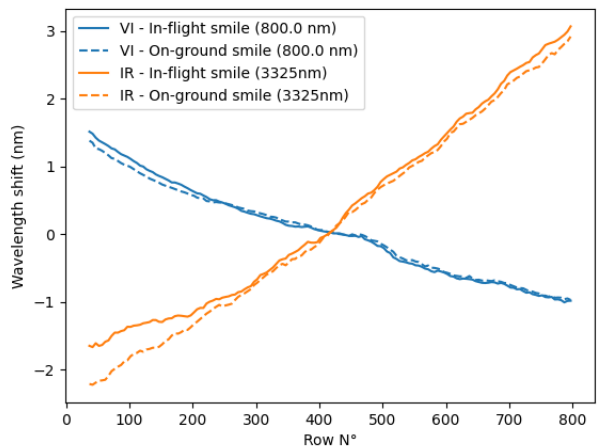

Fig. 28. Comparison of the smile characterization using ICU signatures for both channels before and after launch.

## VI. SUMMARY AND CONCLUSIONS

In this article, we have analyzed ground and flight calibration data to characterize the MAJIS spectral behavior. Specifically, the spectral performances of the instrument were derived in terms of:

- Absolute calibration
- Spectral response

Furthermore, distortions were characterized, and established corrections were provided for:

- the smile (2D maps for the two channels)
- the exploration of the OH temperature (no correction for the VISNIR channel; correction for + 2.1 nm ± 0.2 nm / 10 K for the IR channel)

The spectral performances of the instrument are summarized in Table XI. The NECP data has indicated that the absolute spectral calibration has been shifted after launch by -3.8 nm and -5.4 nm for the VISNIR and IR channels, respectively. Meanwhile, the spectral response and smile have not changed significantly after the launch.

Table XI. Spectral performances (nominal thermal case) against requirements summary

| Criteria | VISNIR Requirements (nm) | VISNIR Performances (nm) | IR Requirements (nm) | IR Performances (nm) | Comments |
|---|---|---|---|---|---|
| **Spectral Range before launch** | 500-2350 | 489-2349 (Option 1) 491-2358 (Option 2) | 2250-5540 | 2268-5556 2271-5555 | |
| **Spectral Range after launch** | 500-2350 | 493-2353 | 2250-5540 | 2273-5561 | |
| **FWHM (oversampling)** | | 2.9/4.6 | | 5.5/7 | No significant change was observed in-flight |
| **FWHM (nominal spectral binning)** | <5.5 | 3.5/5.6 | <10 | 6.6/8.5 | |
| **Spectral Sampling (oversampling)** | | 1.79/1.86 | | 3.0/3.48 | No significant change was observed in-flight |
| **Spectral Sampling (nominal spectral binning)** | 3.65 ± 0.25 | 3.58/3.72 | 6.5 ± 0.60 | 6.0/6.96 | |
| **Smile** | <3.65 | -1.2/2.1 | <7.10 | -1.9/3.1 | A slight distortion decrease was observed in-flight (IR) |
| **Thermal Shift (CWL)** | | - | | +1.8/10K | |

In addition, all relevant spectral performances were adequately characterized for both channels and all binning modes used in-flight (oversampling, nominal, binned×2, and binned×4). Although distortions in terms of smile and OH temperature were observed, corrections were derived to counteract their effects on flight acquisitions. The results demonstrate that the imaging spectrometer meets spectral resolution, range, and smile requirements within the expected OH thermal range. This on-ground spectral calibration should allow the MAJIS scientific team to rapidly and accurately analyze the data from Jupiter and its Galilean Moons.

## VII. AUTHOR DECLARATION

### A. Acknowledgments

French contribution to MAJIS has been technically supported and funded by CNES – CONTRACT CNES – CNRS n° 180 117. Italian contribution to MAJIS has been coordinated and funded by the Italian Space Agency – CONTRACT N. 2021-18-I.0 and supported by the ASI-INAF agreement n. 2023-6-HH.0.

### B. Conflict of interest

The authors have no conflicts to disclose.

### C. Author Contributions

**Alessandra Barbis**: formal analysis (supporting); investigation (equal); resources (equal). **John Carter**: conceptualization (equal); funding acquisition (equal); investigation (equal); resources (equal). **Cydalise Dumesnil**: data curation (equal); project administration (equal); resources (equal). **Gianrico Filacchione**: conceptualization (equal); formal analysis (supporting); methodology (supporting); supervision (supporting); writing – review editing (supporting). **Pierre Guiot**: resources (equal); investigation (equal). **Paolo Haffoud**: formal analysis (lead); investigation (equal); methodology (lead); writing – original draft (lead); writing – review editing (lead). **Yves Langevin**: conceptualization (equal); formal analysis (supporting); funding acquisition (equal); methodology (supporting); supervision (supporting); writing – review editing (supporting). **Benoit Lecomte**: resources (equal); investigation (equal). **Giuseppe Piccioni**: conceptualization (equal); funding acquisition (equal); supervision (supporting). **Cédric Pilorget**: investigation (equal); resources (equal); writing – review editing (supporting). **François Poulet**: conceptualization (equal); data curation (equal); funding acquisition (equal); methodology (supporting); project administration (equal); supervision (lead); writing – original draft (supporting); writing – review editing (lead). **Sébastien Rodriguez**: investigation (equal); methodology (supporting). **Stefani Stefania**: investigation (equal); resources (equal). **Leonardo Tommasi**: conceptualization (equal); investigation (equal); resources (equal); supervision (supporting). **Federico Tosi**: conceptualization (equal). **Mathieu Vincendon**: conceptualization (equal); data curation (equal); methodology (supporting); project administration (equal); supervision (lead); writing – original draft (supporting); writing – review editing (lead).

### D. Data availability

The data that support the findings of this study are available from the corresponding author upon reasonable request.